\documentclass[5p,times,twocolumn]{elsarticle}

\usepackage{amssymb}
\usepackage{amsmath}

\usepackage{color}    
\usepackage{bm}
\usepackage{subfigure}
\usepackage{hyperref}
\usepackage{listings}

\newcommand{\beq}{\begin{equation}}
\newcommand{\eeq}{\end{equation}}
\newcommand{\bea}{\begin{eqnarray}}
\newcommand{\eea}{\end{eqnarray}}

\renewcommand{\vec}[1]{\mathbf{#1}}

\begin{document}
\title{\textit{HEART}: A New X-Ray Tracing Code for Mosaic Crystal Spectrometers}

\author[1,2]{Thomas~Gawne}
\ead{t.gawne@hzdr.de}

\author[1,2]{Sebastian~Schwalbe}

\author[1,2]{Thomas~Chuna}

\author[1,2]{Uwe~Hernandez~Acosta}

\author[3]{Thomas~R.~Preston}

\author[1,2]{Tobias~Dornheim}

\affiliation[1]{organization={Center for Advanced Systems Understanding (CASUS)},
            city={D-02826 Görlitz},
            country={Germany}}
\affiliation[2]{organization={Helmholtz-Zentrum Dresden-Rossendorf (HZDR)},
            city={D-01328 Dresden},
            country={Germany}}
\affiliation[3]{organization={European XFEL},
            city={D-22869 Schenefeld},
            country={Germany}}

\begin{abstract}
We introduce a new open-source Python x-ray tracing code for modelling Bragg diffracting mosaic crystal spectrometers: \emph{High Energy Applications Ray Tracer (HEART)}.
\textit{HEART}'s high modularity enables customizable workflows as well as efficient development of novel features. 
Utilizing Numba's just-in-time (JIT) compiler and the message-passing interface (MPI) allows running \textit{HEART} in parallel leading to excellent performance.
\textit{HEART} is intended to be used for modelling x-ray spectra as they would be seen in experiments that measure x-ray spectroscopy with a mosaic crystal spectrometer.
This enables the user to make predictions about what will be seen on a detector in experiment, perform optimizations on the design of the spectrometer setup, or to study the effect of the spectrometer on measured spectra. However, the code certainly has further uses beyond these example use cases.
Here, we discuss the physical model used in the code, and explore a number of different mosaic distribution functions, intrinsic rocking curves, and sampling approaches which are available to the user. Finally, we demonstrate its strong predictive capability in comparison to spectroscopic data collected at the European XFEL in Germany. \\


\noindent \textbf{PROGRAM SUMMARY}\\
\begin{small}
\noindent
{\em Program Title: High Energy Applications Ray Tracer (HEART)} \\
{\em Developer's repository link:} https://gitlab.com/heart-ray-tracing/HEART  \\
{\em Licensing provisions:} GPLv3  \\
{\em Programming language:} Python $\ge3.10$ \\
{\em Nature of problem:}\\
Mosaic crystal spectrometers are widely-used at high energy density (HED) facilities owing to their very high integrated reflectivities. However, the mosaic nature of the crystal introduces a lot of complexity into the instrument functions of these spectrometers. Understanding how the mosaic crystal will impact the measured spectrum is vital for reliably inferring conditions measured via x-ray spectroscopy and for planning experiments. \\
{\em Solution method:}\\
We have developed a ray tracing code with specific support for mosaic crystals. With the implemented precise dynamical theory models, our ray tracing code simulates accurate detector images enabling realistic comparisons with experiments. 
It takes advantage of the inherent randomness of mosaic crystals to run Monte Carlo simulations of rays passing through the crystal. This also means the detector images produced contain similar photon counting noise that would appear in experiments. A number of options for crystal materials, geometries, mosaic distribution functions, and rocking curves are supported. Effects such as absorption and multiple ray reflections are also treated explicitly.
\end{small}

\end{abstract}

\maketitle


\section{Introduction}\label{sec:intro}

Over the last several decades, x-ray spectroscopy has become an important method for studying and diagnosing systems in a wide range of research scenarios, from material sciences and development~\cite{Kraus2016-hc,Miao2020-ew}, to the study of high energy density systems in the laboratory~\cite{Kritcher2020,Tilo_Nature_2023,Gain-2024}. Due to the long penetration depths of x-rays, they are able to probe the bulk properties of systems, and the high energies of x-ray photons allow them to resonantly interact with the energy levels of atoms.
Additionally, very short x-ray pulses can be produced with a wide range of pulse lengths, from the nanosecond down to the attosecond timescale, allowing for the probing of properties over different characteristic timescales.
Consequently, numerous x-ray spectroscopic techniques have been developed to study different aspects of the electronic and ionic structure of materials, such as x-ray diffraction (XRD)~\cite{Kraus2017,Gorman_2024_Shock}, x-ray emission spectroscopy (XES)~\cite{Vinko2012-fc,Ciricosta_2012_IPD,Ciricosta_2016_IPD}, x-ray absorption spectroscopy (XAS)~\cite{Harmand_2015_XAS,Harmand_2023_XAS}, resonant inelastic x-ray scattering (RIXS)~\cite{Humphries_2020_RIXS, Forte_2024_RIXS}, and x-ray Thomson scattering (XRTS)~\cite{Tilo_Nature_2023,Glenzer_RMP_2009,kraus_xrts}.

When the spectrum of emitted or scattered photons emerge from the target, the light is typically polychromatic in all directions, so in order to measure the spectrum the photons need to be separated in energy. A common way to do this in spectrometers is to use a dispersive crystal which has an appropriate lattice spacing for the photon energies under investigation. For the work here, we focus on Bragg geometry, where photons are dispersed in reflection, as opposed to Laue geometry where the dispersion appears in the transmission through the crystal.
In kinematic diffraction theory, the relationship between a photon's incoming wave vector $\vec{k}$, scattered wave vector $\vec{k}_s$, and the reciprocal lattice with Miller indices $H=(hkl)$ is given by the Laue vector equations~\cite{zachariasen1994theory}:
\begin{equation}
    \vec{k}_s - \vec{k} = \vec{B}_H = h\vec{b}_1 + k\vec{b}_2 + l\vec{b}_3 \, ,
\end{equation}
where $\vec{b}_i$ are the reciprocal lattice vectors. The magnitude of the reciprocal lattice vector $\vec{B}_H$ is then $1/d_{H}$, where $d_{H}$ is the distance between two lattice planes.
Assuming the scattering is elastic -- i.e. $|\vec{k}_s| = |\vec{k}| = hc/E$, where $E$ is the photon energy, $h$ is the Planck constant, and $c$ is the speed of light -- the relationship between the photon energy and the particular angle of incidence that will result in scattering $\alpha_B$ (the Bragg angle) is given by Bragg's law:
\begin{equation}
    \frac{hc}{E} = 2d_{H}\sin(\alpha_B) \, .
    \label{eq:BraggLaw}
\end{equation}
The crystal therefore separates different photon energies in angle. By relating $\alpha_B$ to the position a photon of energy $E$ hits a detector, a spectrum can be measured as photon energy versus number of photons/intensity.

While kinematic theory is useful for providing relatively simple relationships for describing scattering behaviour, it belies the full complexity of the diffraction of x-rays in crystals, which is described by dynamic diffraction theory~\cite{zachariasen1994theory}. Essentially, when treating the incident beam as a wave propagating through the crystal, the incident beam will still be (partially) reflected off the lattice planes at any angle of incidence, and the interference between the incident and diffracting beams inside the crystal result in a finite reflectivity. The reflected intensity versus angle of incidence is a rich diffraction pattern known as the intrinsic rocking curve (IRC) of the crystal. This means the Bragg condition does not need to be exactly satisfied for diffraction to occur, so there is no longer a strict one-to-one relationship between the energy of the photon and its position on the detector. However, since the IRC for thick crystals tends to be very narrow, the relationship appears to hold sufficiently well for many applications.
Furthermore, the integrated reflectivity from dynamic theory will reproduce the kinematic result.
More generally, the relationship between photon energy and position on the detector breaks down due to a number of additional effects beyond the IRC, such as the depth broadening in the crystal and its mosaicity, as well as external effects such as the finite-sized source profiles or the detector response to the incident photons.
As a result, spectral features appear broader than they actually are when measured on a spectrometer, and the broadening effect is called the instrument function. It is ultimately the modelling of the instrument function that we are trying to achieve here.

\begin{figure*}
    \centering
    \includegraphics[width=0.95\textwidth,keepaspectratio]{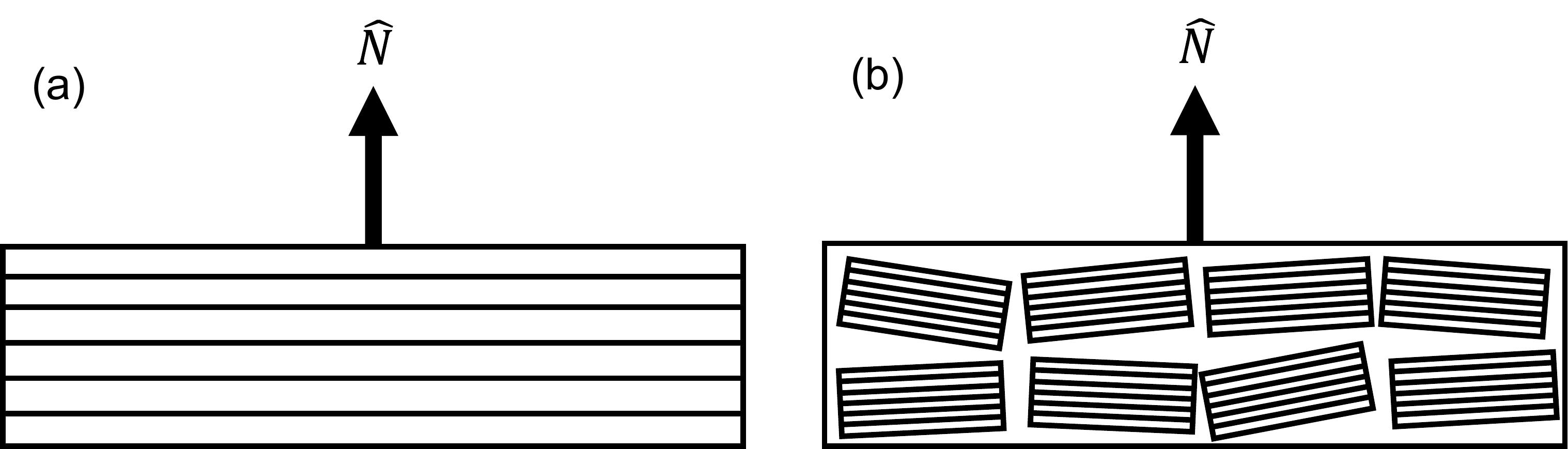}
    \caption{(a) A diagram of a perfect crystal, with its uniform layers of atoms. (b) A diagram of a mosaic crystal, consisting of perfect crystallites which are randomly oriented about the crystal surface normal $\hat{N}$.}
    \label{fig:PerfectMosaic}
\end{figure*}

Returning to the applications of dispersive crystals in the laboratory, we now focus on a particular type: mosaic crystals~\cite{zachariasen1994theory}, which is shown diagrammatically in Fig.~\ref{fig:PerfectMosaic}. In contrast to perfect crystals, which consist of uniform layers of atoms with a well-defined normal, mosaic crystals consist of small perfect crystallites whose normals are randomly oriented to the surface normal of the crystal, with their distribution determined by the crystal's mosaic distribution function (MDF). Whereas in a perfect crystal a photon will only reflect in a small region of the crystal determined by the IRC, in a mosaic crystal there is a probability to find a crystallite with the appropriate angle to satisfy the Bragg condition anywhere in the crystal. The integrated reflectivity of mosaic crystals is therefore substantially higher than that of perfect crystals, making them well-suited to use in situations where there are concerns about the ability to collect a sufficient number of photons to measure a useful signal.

One such situation is in diagnosis of HED systems produced in the laboratory. These systems are characterised by their high temperatures, densities, and pressures~\cite{wdm_book,drake2018high,siegfried_review}; and the conditions can only be produced and maintained for relatively short period of times. Additionally, the breakdown of these conditions is often due to the disassembly of the target, meaning the sample is destroyed. For both reasons, x-ray spectroscopy is used to diagnose the target conditions in the duration they are maintained. For some methods, such as XRD and XES, the cross-sections involved are sufficiently large that enough photons can be collected in single shots. However, other techniques such as XRTS and RIXS have very small cross-sections, so the number of photons that scatter via these processes is very low. Restrictions such as the availability of targets, the repetition rate of the experimental campaign, and the shot-to-shot reproducibility of conditions means that collecting as many photons per shot as possible is highly desirable. For this purpose, spectrometers using mosaic crystals are often employed at HED facilities such as the HED Scientific Instrument at the European XFEL~\cite{Preston_JoI_2020,Zastrau_2021}, the MEC Endstation at the Linac Coherent Light Source (LCLS)~\cite{glenzer2016matter}, and at the National Ignition Facility (NIF)~\cite{Tilo_Nature_2023,Doeppner_2014_MACS,MacDonald_2023_CPS}.

While mosaic crystals have very high integrated reflectivities, this comes at the cost of resolution since the mosaicity also broadens the spectrum~\cite{Preston_JoI_2020,Pak_RoSI_2004,2012_Zastrau_Focal}. Furthermore, the mosaicity also enhances the effective depth a photon can travel in the crystal, which further enhances both the reflectivity and the depth broadening of the crystal~\cite{Schlesiger_JAC_2017,Gawne_2024_Effects}.
Consequently, the instrument function of mosaic crystals is in general very complex, varies in shape across the crystal, and depends on the specific position of the spectrometer relative to the source~\cite{Gawne_2024_Effects}.
For many spectroscopic techniques, such as RIXS and XRTS, the probed structure of the system is encoded in the shape of the spectrum. As a result, to get useful information about the system, the instrument function needs to be properly taken into account, but for mosaic crystals this is non-trivial. In general, accurately measuring the full instrument function across the spectrometer for each experiment is unfeasible, so modelling the instrument function can be a useful alternative. Furthermore, the ability to accurately model the instrument function can be useful in removing the effect of the instrument function via forward-modelling approaches, or in planning experiments to see what effect the instrument function will have on the expected measured spectrum.

The construction of accurate analytical models for mosaic crystals is challenging due to the complex way that the various broadening effects interact with one another, such that they cannot necessarily be cleanly separated. However, an alternative approach is to take advantage of the statistical nature of the distribution of the crystallites. When an x-ray beam is incident on a mosaic crystal, the random distribution of the crystallites means that the scattered photons can be treated as no longer being coherent with one another. Therefore, rather than treating the diffraction of the beam, the diffraction of individual photons can be treated, which just involves tracing the path each individual photon would take. Due to the statistical distribution of the crystallites and the large number of rays involved in producing realistic spectra, Monte Carlo ray tracing simulations are well-suited to modelling the instrument function of a mosaic crystal.
Monte Carlo simulations are a well-established class of simulation methods when interactions are probabilistic, allowing for efficient generation of predictions by taking advantage of the random nature of events. In the context of radiation and particle transport through matter, there are a number of popular Monte Carlo codes such as \textit{PENELOPE}~\cite{PENELOPE_1995,PENELOPE_2018}, \textit{EGSnrc}~\cite{egsnrc}, and \textit{Geant4}~\cite{Geant4_2003,Geant4_2006,Geant4_2016}.
For ray tracing, Monte Carlo methods are also popular for generating spectra~\cite{2012_Zastrau_Focal,SHADOW3,Smid_CPC_2021,voxTrace_2023} since reflectivity, absorption, and transmission curves also represent the probability distributions for these interactions occurring.
Furthermore, Monte Carlo ray tracing benefits from being able to simply continue running simulations until the desired signal-to-noise ratio (SNR) is reached on the modelled detector since the interaction events are treated independently from each other. In fact, the statistical nature of ray tracing simulations is well-suited for modelling experiments since photon measurements are discrete events, so the resulting noise in a ray tracing simulation can be representative of the noise seen in experimental data.

There are a number of ray tracing codes available. One of the most popular codes is \textit{SHADOW}~\cite{SHADOW, SHADOW3, SHADOW4}. \textit{SHADOW} is a very general purpose ray tracing and diffraction code, allowing the simulations of numerous different types optical elements (not just crystals), and its results have been well-tested and verified over several decades. Its model for mosaic crystals is based directly on the mosaic crystal diffraction theory of Ref.~\cite{zachariasen1994theory}. This also means means it is restricted to flat mosaic crystals, and only those with a Gaussian MDF~\cite{delRio_1992_Conceptual, delRio_2013_Mosaic}. However, many spectrometers at x-ray facilities utilise bent mosaic crystals, some with distributions that are different from a Gaussian, such as widely-used highly annealed pyrolytic graphite (HAPG) which has a Lorentzian MDF~\cite{Gerlach_JAC_2015}.

More recently the open source ray tracing code \textit{MMPXRT} was released~\cite{Smid_CPC_2021}, which was built to perform ray tracing simulations for spectrometers containing a mosaic crystal, with support for flat, cylindrical, and toroidal geometry. Since then, \textit{MMPXRT} has seen some use in the modelling spectrometers for experiments~\cite{Sanders_2021_Microstructured, Pan_2023_Imaging, Hesselbach_2024_Platform, Kumar_2024_High, Kettle_2024_Extended}.
However, on examining \textit{MMPXRT}, we have identified a number of areas for improvement.
First, the current version of \textit{MMPXRT} only supports simulations of mosaic crystals with a Lorentzian MDF: while this may be fine for modelling crystals such as HAPG, it is not sufficiently accurate for modelling other commonplace crystals with alternate MDFs, such as highly oriented pyrolytic graphite (HOPG) which has a Gaussian MDF~\cite{Gerlach_JAC_2015}. Furthermore, the IRC is modelled as a box function in \textit{MMPXRT}; but, as will be shown later in this work, such a model does not produce sufficiently accurate instrument functions since the IRC has heavy tails. Additionally, the reflection model in \textit{MMPXRT} neglects absorption effects and multiple reflections within the crystal, both of which have a noticeable effect on the (integrated) reflectivity~\cite{Schlesiger_JAC_2017} and shape of the crystal instrument function~\cite{Gawne_2024_Effects}.

In order to address these shortcomings in \textit{MMPXRT}, as well as wishing to develop a flexible platform for future studies, we have developed a new x-ray Monte Carlo mosaic crystal ray tracing code -- the \emph{High Energy Applications Ray Tracer (HEART)} -- which we introduce here.
\textit{HEART} is an open source x-ray ray tracing code written in Python 3.
The code utilises Numba's~\cite{Numba} just-in-time (JIT) compiler to enhance performance and enable multithread parallelization.
Additionally,  utilizing the mpi4py package~\cite{mpi4py}, HEART can run parallel on multiple nodes using MPI.
The code is very flexible, able to run simulations with a predefined distribution, or accepting a fully custom ray set of wave vectors, photon energies, and polarizations.
Multiple reflections are accounted for directly by tracing the rays' full paths until they are either absorbed in the crystal or they leave.
The code uses the \textit{xraylib}~\cite{Brunetti_2004_xraylib,Schoonjans_2011_Xraylib} and \textit{XrayDB}~\cite{xraydb} libraries to provide absorption and structure factor data for the different elements, allowing any material to be considered in the code.
Additionally, as \textit{HEART} is a Monte Carlo ray tracer, the detector images produced by \textit{HEART} are simply a count of the number of photons that land in each pixel. Therefore, the photon counting noise inherent in experiments develops naturally in \textit{HEART}.
We therefore expect that \textit{HEART} will be useful for a number of applications, from planning experiments, to performing theoretical studies with controlled computer experiments.

This article is structured as follows: first, we describe in detail the underlying model used in \textit{HEART}; we then outline the usage of the code; next, we present some results from ray tracing simulations using the code, which demonstrates the code's ability to make good predictions of experiments, and to compare different models in the code; and finally we summarise the article's main findings.


\section{Physical Model}\label{sec:model}

In the current version of \textit{HEART}, there are three components: the source of the photons, the mosaic crystal, and the detector. In this section, the treatment of each component is discussed, though it is worth noting that it is the crystal which has the most detailed modelling in the code. 

\subsection{Photon Source}

The source is just the set of origins $\{\bm{o}_s\}$, photon energies $\{E\}$, initial unit wave vectors $\{\bm{\hat{k}}_0\}$ and, optionally, initial unit polarization vectors $\{\bm{\hat{p}}_0\}$ of the initialised rays.
\textit{HEART} is flexible on the input: a user can either define an intensity profile that will be used as a probability distribution function (PDF) to sample the photons energies, or it can accept an array of photon energies to trace instead. Additionally, the outgoing directions of the rays and their origins can either be provided by the user or left to the code to sample. Finally, polarization effects in the crystal can be considered if a user provides an array of initial polarizations for each of the rays.

\subsection{The Mosaic Crystal}

\begin{figure*}
    \centering
    \includegraphics[width=0.95\textwidth,keepaspectratio]{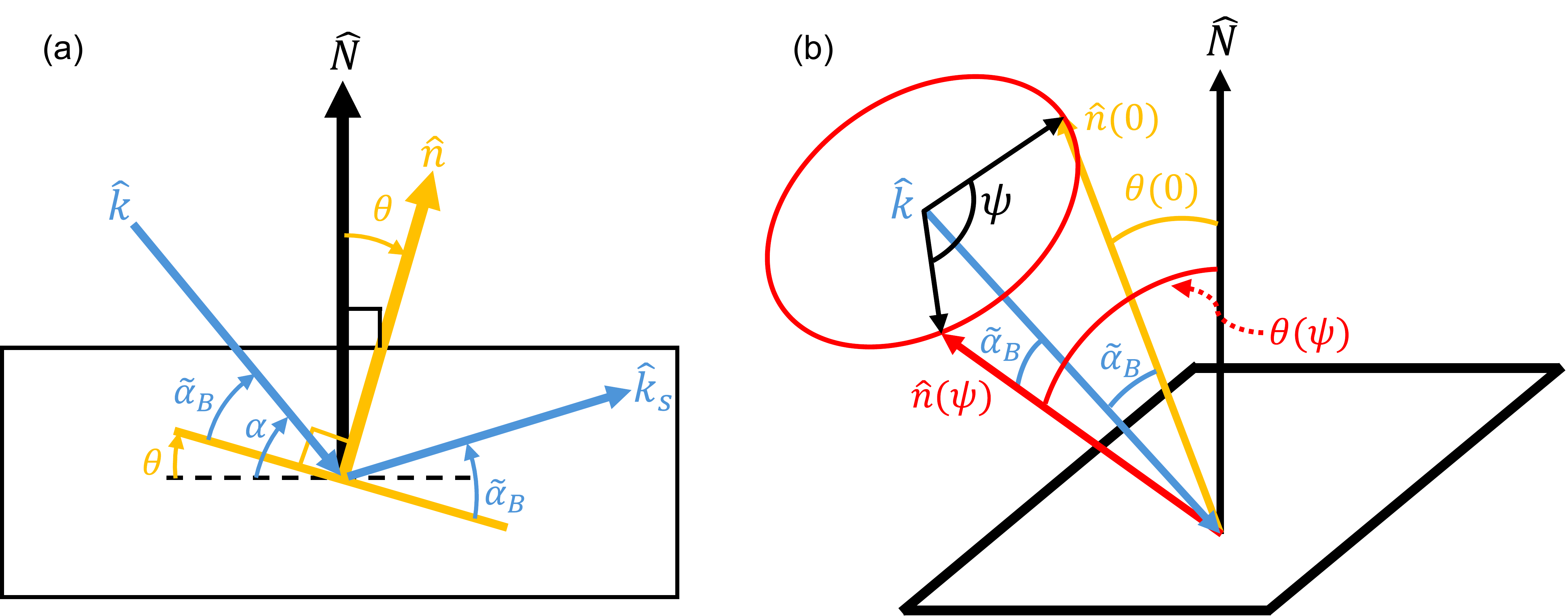}
    \caption{Diagram of the scattering geometry in mosaic crystals. (a) A view of the scattering geometry in the plane containing the crystal surface normal vector $\hat{N}$ and the incident unit wave vector of the ray $\hat{k}$. The angle of incidence of the ray $\alpha$ is defined as $\sin(\alpha) = -\hat{k} \cdot \hat{N}$. The crystallite angle $\theta$ is defined as $\cos(\theta) = \hat{N} \cdot \hat{n}$, where $\hat{n}$ is the normal vector of the crystallite.
    The angle $\tilde{\alpha}_B = \alpha_B + \Delta$ is the angle of incidence of the photon on the crystallite, where $\Delta$ is the rocking curve angle, and $\Delta=0$ corresponds to the Bragg condition being satisfied on the crystallite.
    In the planar solution is $\theta = \alpha - \tilde{\alpha}_B$. $\hat{k}_s$ is the unit wave vector of the scattered photon, which is still at angle an $\tilde{\alpha}_B$ to the crystallite, and an angle $2\tilde{\alpha}_B - \alpha$ to the crystal surface normal.
    (b) The full set of crystallites that the photon can be incident on at an angle $\tilde{\alpha}_B$ is a circle around $\hat{k}$ (indicated by the red circle), so the angle $\theta \equiv \theta(\psi)$, where $\psi$ is the angle of the crystallite vector on the circle, with $\psi=0$ corresponding to the smallest angle $\theta(0) = |\alpha - \tilde{\alpha}_B|$. The maximum angle is a second planar solution $\theta(\pi) = |\pi - \alpha - \tilde{\alpha}_B|$.
    }
    \label{fig:ScatterGeom}
\end{figure*}

In the current version of the code, only a single crystal can be added to the spectrometer. However, this is quite common, particularly in HED facilities, so it is not much of a restriction.
The crystal is a class (\texttt{Crystal}) that contains the physical properties of the crystal -- the MDF, the IRC, the atoms in the crystal, their position in the unit cell, the unit cell volume, the Miller indices $H$ of the scattering plane, and the crystal temperature and Debye temperature. These properties are used to determine the reflection and absorption properties of the crystal.

Additionally, there are subclasses from \texttt{Crystal} for specific shapes of the crystal. Currently, \textit{HEART} supports flat crystals (\texttt{FlatCrystal}) and cylindrical crystals (\texttt{CylinderCrystal}). These subclasses contain information about the position of the crystal in space and the spatial extent of the crystal, which are used to determine if a ray will initially intersect a crystal, and to determine if a ray's current position is inside the crystal or not. These are separate functions since the latter can be evaluated much more quickly than the former, and the latter may be evaluated more often than the former depending on how many times the ray reflects inside the crystal.

For the remainder of this subsection, we focus on the description of the physical properties of the crystal. This geometry described in this section is shown diagrammatically in Fig.~\ref{fig:ScatterGeom}, and the symbols used there are used throughout the following description.

\subsubsection{Intrinsic Rocking Curve (IRC)}

From dynamical diffraction theory~\cite{zachariasen1994theory}, as an incident beam travels through a perfect crystal it can be partially reflected off the lattice planes, resulting in a returning diffraction beam. The incident and diffracted beams are coherent and interfere with one another. The result is that (a) an incident beam does not need to satisfy the Bragg condition in order to diffract off a crystal, and (b) the intensity of the diffracted beam varies in intensity with the angle of incidence of the incident beam.
This diffraction pattern is the rocking curve of the perfect crystal, $R(\Delta, \Lambda)$, where $\Lambda$ are the parameters that characterise the shape of the rocking curve; $\Delta$ is the rocking curve angle, which is the difference between the angle of incidence on the crystal $\alpha$ and the Bragg angle $\alpha_B$.

More specifically, the rocking curve describes the ratio of the intensity of the diffracted beam to the intensity of the incident beam, and in mosaic crystals this also covers the mosaic effects. To distinguish the mosaic effect from the rocking curves of the crystallites, we label the latter the IRC of the crystal.

Integrating the IRC over the angle $\Delta$ provides the integrated reflectivity of the crystallite. However, for the purposes of this work the IRC is normalized to unity by pulling the integrated reflectivity out as a constant. This becomes useful later when the reflectivity of the mosaic crystal is calculated.

The IRC for a crystallite of arbitrary thickness $t_0$ can be calculated exactly using dynamical diffraction theory as described in Ref.~\cite{zachariasen1994theory}, so long as the crystallite surfaces are infinite parallel planes.
For x-rays, the infinite plane approximation is good~\cite{delRio_1992_Conceptual} since the crystallites tend to have a size $\gtrsim$100~nm~\cite{2012_Zastrau_Focal}, and so are much larger than typical x-ray wavelengths ($1.24$~nm for a 1~keV photon).
The second assumption does not necessarily hold if the crystal is bent. Indeed, for bent perfect crystals, the bending of the lattice planes results in an IRC that is substantially more asymmetric and extended versus the plane parallel case, even for bending radii that are several tens of millimetres~\cite{Takagi_1962_Dynamical,Taupin_1964_Theorie,Caciuffo_1987_Monochromators,Uschmann_1993_Xray,Zastrau_2014_Bent}. Furthermore, it is relatively computationally expensive to solve the Takagi-Taupin equations that describe the bent crystal IRC~\cite{Takagi_1962_Dynamical,Taupin_1964_Theorie} within a ray tracing code.
That being said, many mosaic crystals such as HAPG are quite pliable, suggesting the crystallites do not deform or experience significant strain when bent. For this reason, we treat the crystallite surfaces as being parallel.

Within \textit{HEART}, the IRC can be calculated for a single crystallite thickness using the equations derived in Ref.~\cite{zachariasen1994theory}.
However, the measured IRC in HAPG closely resembles a Voigt profile since it is the average the IRC over multiple crystallite thicknesses~\cite{Gawne_2024_Effects,Gerlach_JAC_2015, Legall_High_2006}. For this reason, a pseudo-Voigt profile has been implemented as the preferred option for modelling the IRC, where the input are then the full widths at half maxima (FWHM) of the Gaussian and Lorentzian components; i.e $\Lambda=(\Lambda_G, \Lambda_L)$.
We use a weighted Gaussian-Lorentzian sum for the pseudo-Voigt profile of the form~\cite{Thompson_1987_Ritveld, Ida_2000_Extended}:
\begin{equation}
    R(\Delta, \Lambda_G, \Lambda_L) = \eta L(\Delta, \Lambda) + (1 - \eta) G(\Delta, \Lambda) \, ,
\end{equation}
where
\begin{equation}
\begin{split}
   \Lambda \equiv \Lambda(\Lambda_G, \Lambda_L) &= \left( \Lambda_G^5 + 2.69269 \Lambda_G^4 \Lambda_L + 2.42843 \Lambda_G^3 \Lambda_L^2 \right. \\ 
   &\left. + 4.47163 \Lambda_G^2 \Lambda_L^3 + 0.07842 \Lambda_G \Lambda_L^4 + \Lambda_L^5 \right)^{1/5} \, ,
\end{split}
\end{equation}
\begin{equation}
    \begin{split}
        \eta &\equiv \eta(\Lambda_G, \Lambda_L) \\
        &= 1.36603\left(\frac{\Lambda_L}{\Lambda}\right) - 0.47719\left(\frac{\Lambda_L}{\Lambda}\right)^2 + 0.11116\left(\frac{\Lambda_L}{\Lambda}\right)^3 \, ,
    \end{split}
\end{equation}
\begin{align}
    G(\Delta, \Lambda) &= \sqrt{\frac{4\ln2}{\pi \Lambda^2}} \exp\left(-4\ln2\frac{\Delta^2}{\Lambda^2}\right) \, , \label{eq:Gaussian} \\
    L(\Delta, \Lambda) &= \frac{\Lambda}{2\pi} \frac{1}{ (\Lambda/2)^2+ \Delta^2} \, . \label{eq:Lorentz}
\end{align}
Additionally, a box function IRC has been implemented as well, although it is unlikely to give accurate results due to the lack of tails in the distribution. Finally, the contribution of the IRC can be disabled altogether to investigate its contribution to the crystal instrument function.

Since the peak position of the IRC $\Delta_M$ does not, in general, occur at $\Delta_M = 0$, the maximum point can be shifted when defining the rocking curve. Alternatively, since $\Delta_M$ depends photon energy, the user can also provide a custom function that relates $\Delta_M$ to the photon energy.
Additionally, the width of the rocking curve also depends on photon energy, so the user can optionally provide a function to define the dependency of the FWHM of the rocking curve. In the case of the Voigt IRC, the code maintains the Lorentzian-ness and Gaussian-ness of the original input widths.

To help a user decide on the rocking curve width and central position, the \texttt{Crystal} class has a method \texttt{Crystal.diffraction\_rocking\_curve()} which can average the dynamical diffraction IRC over a user-defined distribution of crystallite thicknesses, which they can then fit against.

\subsubsection{Mosaic Distribution Function (MDF)}

The defining characteristic of a mosaic crystal is that it consists of small domains of perfect crystallites whose normals are oriented randomly to the surface normal of the crystal.
The distribution of the crystallite normals is given by the mosaic distribution function, $W(\theta, \Gamma)$. It describes the probability of finding a crystallite with a normal vector $\hat{n}$ when the local (surface) normal vector is $\hat{N}$, where the crystallite angle $\theta = \arccos(\hat{n} \cdot \hat{N})$, and $\Gamma$ defines the width of the distribution (see Fig.~\ref{fig:ScatterGeom} for reference).
Assuming the Bragg condition needs to be met in order for a photon to reflect, the random distribution of the crystallites means there is a finite probability that a photon will encounter a crystallite that allows it to satisfy the Bragg condition.

A common form for the MDF is a Gaussian~\cite{zachariasen1994theory}, which is in fact the form of the MDF in HOPG crystals~\cite{Beckhoff_1996_New, Freund_1996_Xray, Gerlach_JAC_2015}. An even simpler uniform distribution has been proposed for simulations before~\cite{Sears_1997_Bragg}. Recent measurements of the MDF in HAPG crystals have revealed that it resembles the sum of two Lorentzians, though ray tracing simulations with a single Lorentzian MDF achieved good visual agreement with experiment~\cite{Gerlach_JAC_2015}. The difference between a Gaussian and a Lorentzian MDF has a substantial impact on predicted spectral line shapes from ray tracing simulations~\cite{Gerlach_JAC_2015}.

More recently, it has been noted that the crystallite normal vectors $\hat{n}$ are distributed on the unit sphere, with $\hat{N}$ as the mean~\cite{Bornemann_2020_Multiple}. This also means the distribution is wrapped, since the crystallite normal orientations are only uniquely defined for $\theta \in [0, \pi]$.
For a Gaussian MDF, assuming the crystallite orientation has isotropic covariance, the distribution is the von Mises-Fisher distribution~\cite{Bornemann_2020_Multiple,Pewsey_2021_Recent}:
\begin{equation}
    \tilde{W}(\hat{n}; \hat{N}, s) = \frac{1}{2 \pi s^2 [1 - \exp(-2/s^2)]} \exp\left(-\frac{(\hat{n} - \hat{N})^2}{2 s^2}\right) \, ,
    \label{eq:vMF}
\end{equation}
where $s^2$ is the variance of the distribution, and $\hat{n} \cdot \hat{n} = \hat{N} \cdot \hat{N} = 1$, since they are unit vectors.
As the lengths of the vectors are constrained, the only free parameters are the crystallite angle $\theta$ and an azimuthal angle $\xi$ about $\hat{N}$. Clearly the MDF in Eq.~(\ref{eq:vMF}) is uniform in $\xi$, and this is assumed to be true throughout this work.
With these restrictions, the MDF can be reduced to a function of $\theta$:
\begin{equation}
    W_{\rm MF}(\theta; s) = \frac{1}{2 \pi s^2 [1 - \exp(-2/s^2)]} \exp\left(\frac{\cos(\theta)-1}{s^2}\right) \, ,
    \label{eq:MF}
\end{equation}
where the normalisation is:
\begin{equation}
    \int_{-\pi}^{\pi} d\psi \int_{0}^{\pi} d\theta \sin(\theta) W_{\rm MF}(\theta; s) = 1 \, .
\end{equation}
Though this function does not immediately resemble a Gaussian, in practice the typically narrow widths of the MDF ($\sim 1^\circ$ for HOPG, $\sim0.1^\circ$ for HAPG) means that its contribution to the reflectivity still resembles a Gaussian of the form in Eq.~(\ref{eq:Gaussian}) (with $\Delta \rightarrow \theta$ and $\Lambda \rightarrow \Gamma = 2s\sqrt{2\ln2}$)~\cite{Bornemann_2020_Multiple}, since $\cos(\theta)-1 \simeq \theta^2$.

In a similar vein, we wish to extend this idea of distributing the crystallite normals on the unit sphere for a Lorentzian distribution. An initial obvious choice would be a spherically-wrapped Lorentzian distribution~\cite{Pewsey_2021_Recent,Shogo_2020_Some}:
\begin{equation}
    W_{\rm SWL}(\theta; \Gamma) = \frac{1}{4\pi} \frac{\sinh^2(\Gamma/2)}{[\cosh(\Gamma/2) - \cos(\theta)]^2} \, ,
    \label{eq:SWL}
\end{equation}
where $\Gamma$ is akin to the FWHM of the Lorentzian distribution.
However, in practice we found this distribution has the crystallite normals that are distributed too densely around the surface normal. This leads to insufficient reflectivity across the crystal away from the nominal Bragg condition, which in turn meant there was insufficient asymmetry~\cite{Gawne_2024_Effects} in simulated spectra compared to experiment.
In fact, this is an issue that we also observed for the standard Lorentzian distribution of the form Eq.~(\ref{eq:Lorentz}).
An alternative form that we propose here is to assume the $\theta^2$ in the Lorentzian should be replaced directly with $(\hat{n} - \hat{N})^2$, giving a PDF of:
\begin{equation}
    W_{\rm WL}(\theta; \Gamma) = \frac{1}{\pi \ln[1 + (4/\Gamma)^2] } \frac{1}{(\Gamma/2)^2 + 2 - 2\cos(\theta)} \, .
    \label{eq:WL}
\end{equation}
As will be shown later, this distribution looks similar to the Lorentzian crystallite distribution, however it results in substantially different reflectivity curves that are much flatter across the crystal, resulting in very asymmetric distributions that appear to agree well with experiment (see Section~\ref{sec:results}).

We now turn our attention to the possible orientations for the crystallites.
For brevity, we assume that a crystallites needs to be oriented such that the Bragg condition is satisfied, however this can be generalized to a generic rocking curve angle by replacing $\alpha_B \rightarrow \alpha_B + \Delta$ in the following.
For the Bragg condition to be satisfied, we require that $\hat{k}\cdot\hat{n}=-\sin(\alpha_B)$.
The simplest solution is that $\hat{n}$ is rotated from $\hat{N}$ by an angle $\theta_0 = \alpha - \alpha_B$ in the plane containing $\hat{k}$ and $\hat{N}$; i.e.

\begin{align}
        \hat{n}_0 &= \mathcal{R}\left(\sec(\alpha)\hat{N} \times \hat{k}, \theta_0  \right)\hat{N} \nonumber  \\
        &= \sec(\alpha) \sin(\theta_0 )\hat{k} + \left[\cos(\theta_0 ) + \tan(\alpha)\sin(\theta_0 )\right]\hat{N} \, ,
\end{align}
where $\mathcal{R}(\hat{v}, \theta)$ is the rotation matrix for a rotation of angle $\theta$ about the unit vector $\hat{v}$, $\sec(\alpha)$ is the normalization of $\hat{N} \times \hat{k}$, and in the last line the crystallite vector $\hat{n}_0$ is expressed in terms of the unit vectors $\hat{k}$ and $\hat{N}$ by using the Rodrigues rotation formula.

More generally, the Bragg condition can clearly be satisfied by any crystallite at angle $\alpha_B$ from the wave vector. This means that possible crystallite normals lie on a circle around $\hat{k}$ (see Fig.~\ref{fig:ScatterGeom}).
This family of solutions can be constructed by rotating the initial crystallite vector $\hat{n}_0$ about the ray vector by an arbitrary angle $\psi$ to give a set of crystallite vector $\hat{n}_\psi$:
\begin{align}
    \hat{n}(\psi) &= \mathcal{R}(\hat{k}, \psi)\hat{n}_0 \nonumber\\
    &=\cos(\psi)\hat{n}_0 + \sin(\psi)(\hat{k}\times\hat{n}_0) \\ 
    &\phantom{=}- \sin(\alpha_B)[1-\cos(\psi)]\hat{k} \, \nonumber .
\end{align}
For completeness, the crystallite vectors that will satisfy the Bragg condition can be written in terms of $\hat{k}$ and $\hat{N}$ as:
\begin{equation}
    \begin{split}
        \hat{n}(\psi) &= \left[\cos(\psi)\sin(\theta_0 )\sec(\alpha) - \sin(\alpha_B)\left\{1 -  \cos(\psi)\right\} \right] \hat{k} \\
        &+ \cos(\psi)\left[ \cos(\theta_0 ) + \tan(\alpha) \sin(\theta_0 ) \right] \hat{N} \\
        &+ \sin(\psi)\left[ \cos(\theta_0 ) + \tan(\alpha) \sin(\theta_0 ) \right] (\hat{k}\times\hat{N}) \, .
    \end{split}
    \label{eq:CrystallitePsi}
\end{equation}
In order to determine the probability of reflecting from a specific crystallite $\hat{n}_\psi$, we need the crystallite angle $\theta(\alpha, \psi, \alpha_B) = \arccos(\hat{n}_\psi \cdot \hat{N})$.
From Eq.~(\ref{eq:CrystallitePsi}), and using trigonometric identities, the angle $\theta(\alpha, \psi, \alpha_B)$ is related to $\psi$ by~\cite{delRio_1992_Conceptual,delRio_2013_Mosaic}:
\begin{equation}
    \begin{split}
    \theta(\alpha, \psi, \alpha_B) = \arccos[&\cos(\psi)\cos(\alpha)\cos(\alpha_B) \\
    &+\sin(\alpha)\sin(\alpha_B)] \, .
    \end{split}
    \label{eq:ThetaPsi}
\end{equation}
To orient the crystallite normal then, first the crystal normal is rotated by an angle $\alpha-\alpha_B$ about the axis $\hat{N}\times\hat{k}$, then this new normal is rotated about the ray axis by an angle $\psi$.

From Eq.~(\ref{eq:ThetaPsi}), it is easy to see that for $\psi=0$, $\theta_0=|\alpha-\alpha_B|$, which was the starting point.
Additionally, there is a second solution in the $\hat{k}$--$\hat{N}$ plane for $\psi=\pm \pi$, $\theta_\pi=|\pi - \alpha - \alpha_B|$.
Evidently, $\psi=0$ and $\psi=\pm\pi$ represent the limits on the size of the angle $\theta$, with the former being the smallest possible angle $\theta$ and therefore the most probable from the MDF, and the latter being the largest angle and the least probable.

As will be shown later, the reflectivity cross-section involves an integral over the angle $\psi$, not $\theta$. Therefore, in determining the orientation of the crystallite that the photon actually reflects from, it is the angle $\psi$ that needs to be determined via random sampling, with the angle $\theta$ inferred from $\psi$ (see \ref{sec:ExtraSampling} for more information).

\subsubsection{Absorption and Reflection}\label{sec:ReflModel}

\begin{figure*}
    \centering
    \includegraphics[width=\textwidth,keepaspectratio]{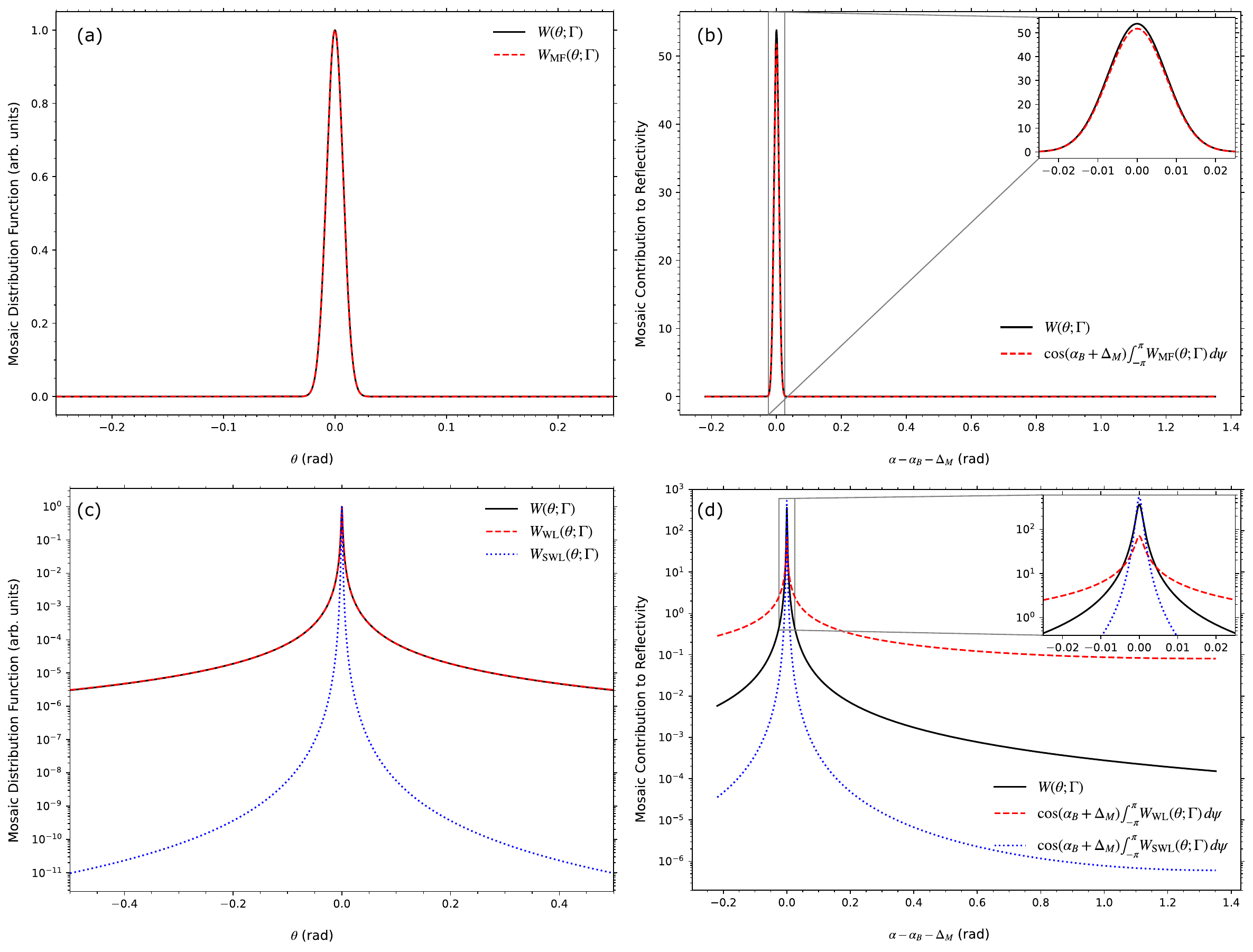}
    \caption{(a) A comparison of the shapes of the MDF for a Gaussian $W(\theta, \Gamma)$ (Eq.~(\ref{eq:Gaussian})), and the von Mises-Fisher distribution $W_{\rm MF}(\theta, \Gamma)$ (Eq.~(\ref{eq:MF})), where $\Gamma=1.0^{\circ}$ is the FWHM of the Gaussian. The distributions are normalized to their peaks to compare the shapes.
    (b) A comparison of the reflectivity curves for the distributions in (a) for $\alpha_B+\Delta_M = 12.55^\circ$, using Eq.~(\ref{eq:SimpleZachXsection}) for $W(\theta, \Gamma)$, and Eq.~(\ref{eq:SimpleReflXsection}) for $W_{\rm MF}(\theta, \Gamma)$. Both the MDFs and reflectivity curves have very similar shapes.
    (c) A comparison of the shapes of the MDF for a Lorentzian $W(\theta, \Gamma)$ (Eq.~(\ref{eq:Lorentz})), $W_{\rm WL}(\theta, \Gamma)$ (Eq.~(\ref{eq:WL})), and $W_{\rm SWL}(\theta, \Gamma)$ (Eq.~(\ref{eq:SWL})), where $\Gamma=0.1^{\circ}$ is the FWHM of the Lorentzian. The distributions are normalized to their peaks to compare the shapes.
    (d) A comparison of the reflectivity curves for the distributions in (c) for $\alpha_B+\Delta_M = 12.55^\circ$, using Eq.~(\ref{eq:SimpleZachXsection}) for $W(\theta, \Gamma)$, and Eq.~(\ref{eq:SimpleReflXsection}) for $W_{\rm WL}(\theta, \Gamma)$ and $W_{\rm SWL}(\theta, \Gamma)$. Note that while $W$ and $W_{\rm WL}$ have similar MDFs, their reflectivity curves are substantially different.
    }
    \label{fig:Reflectivities}
\end{figure*}

The first check performed by the code is whether a ray leaving the source intersects the crystal. If it does not, then the ray is immediately discarded. If the ray does intersect the crystal, then the ray's path through the crystal from the surface will be traced.
In \textit{HEART}, multiple reflections are considered explicitly -- that is, a photon is not just traced to see where an event may occur in the crystal before allowing it to leave, but rather its full path is traced until the ray either exits the crystal or is absorbed. For many mosaic crystals, multiple reflections are an important consideration as a considerable number of photons can reflect multiple times before either leaving the crystal or being absorbed~\cite{Schlesiger_JAC_2017}.

The first thing that must calculated is how far the ray would travel before an event -- either absorption or reflection -- occurs. The probability of an absorption or reflection event occurring is then given by sampling an exponential probability distribution~\cite{zachariasen1994theory}:
\begin{equation}
    P(d) \propto \exp(- [\mu_0 + \gamma_0 \sigma]l) \, ,
    \label{eq:AnyEvent}
\end{equation}
where $\mu_0$ is the linear absorption coefficient for a photon of energy $E$ in the material of the crystal, $\gamma_0 = \sin(\alpha)$, $\sigma$ is the reflectivity coefficient, and $l$ is the path length of the ray.
The energy loss of a beam through the crystal is being described in terms of an effective attenuation coefficient $\mu = \mu_0 + \gamma_0 \sigma$.
Note that the reflection term is scaled by the angle of incidence whereas the absorption term is not. The reason is that the reflection term is formulated in terms of a photon reflecting off a layer of crystal at a depth $z=\gamma_0l$ from the photon's starting point. On the other hand, the absorption is just dependent on the total path length through the crystal $l$.

The absorption term $\mu_0$ is calculated for each photon energy using the \texttt{CS\_Total} function function from the \emph{xraylib} library~\cite{Brunetti_2004_xraylib,Schoonjans_2011_Xraylib}, which provides the mass attenuation coefficients $\mu_0/\rho$ for each of the species in the crystal, where $\rho$ is the density of the crystal. This includes photoabsorption (true absorption), Rayleigh scattering, and Compton scattering.
Here, we treat the two scattering contributions as additional true absorption terms. While they are strictly not, their effect is to randomly scatter photons in space. The scattered rays appear as a background on a detector that is positioned far away from the crystal, and this background gets weaker the further the detector is from the crystal, such that at a large distance from the crystal the scattering resembles true absorption. Treating these effects as true absorption improves the efficiency of the code since these scattering geometries no longer need to be consider, nor does the change in the photon energy in Compton scattering need to be accounted for.

To get a form for the reflection term $\gamma_0 \sigma$, we assume negligible primary extinction (the energy loss due to diffraction through a given crystallite).
Under this assumption, the reflection cross-section for a mosaic crystal is~\cite{zachariasen1994theory}:
\begin{equation}
    \begin{split}
        \gamma_0 \sigma(\alpha, \alpha_B) =& Q  G \int_{-\infty}^{\infty} R(\Delta; \Lambda) W(\alpha-\alpha_B-\Delta; \Gamma) d\Delta  \, ,
    \end{split}
    \label{eq:ZachReflXsec}
\end{equation}
where $\alpha_B+\Delta$ is the angle of incidence of the photon on the crystallite, and
\begin{equation}
    Q = \left| \frac{r_e F_H}{V} \right|^2 \frac{\lambda^3}{\sin(2\alpha_B)} \, ,
\end{equation}
where $F_H$ is the structure factor for the Miller indices $H$, $V$ is the volume of the unit cell, $\lambda$ is the wavelength of the incident photon, and $r_e$ is the classical electron radius. The term $G \equiv G(\hat{p}_i, \hat{k}_s)$ is the polarization term, which depends on the orientation of the electric field of the incident photon $\hat{p}_i$ to the unit wave vector of the scattered photon $\hat{k}_s$:
\begin{equation}
    G = 1 - (\hat{p}_i \cdot \hat{k}_s)^2 = 1 - \sin^2(2\alpha_B)\sin^2(\phi) \, ,
\end{equation}
where $\phi$ is than angle of the polarization vector about the ray, defined here as $\cos(\phi)= \hat{p}_i \cdot (\hat{N}\times\hat{k})\sec(\alpha)$.
For an unpolarized beam, $G$ is averaged over $\phi$ to give:
\begin{equation}
    G_{\rm unpol} = \frac{1 + \cos^2(2\alpha_B)}{2} \, .
\end{equation}
Within \textit{HEART}, the contribution of the Debye-Waller (DW) factor to $F_H$ can be accounted for by setting a Debye temperature $T_D > 0$ when defining the crystal. The calculation of the DW factor follows the approach of Ref.~\cite{Seeger_2002_Mosaic}. 
In addition to the dependence of the reflectivity on the photon energy via the dependence of the Bragg angle in $\gamma_0 \sigma $ and $Q$, $F_H$ also depends directly on the photon energy via the anomalous dispersion~\cite{zachariasen1994theory}. The structure factor for each photon energy is calculated within \textit{HEART} using the \texttt{XrayDB}~\cite{xraydb} library to provide the anomalous dispersion terms for each photon energy and the atomic form factors~\cite{Waasmaier_1995_New, Chantler_2000_Detailed}.

In the case that the IRC is much narrower than the MDF ($\Lambda \ll \Gamma$), which is often the case, then the MDF looks roughly constant across the width of the IRC. Therefore, Eq.~(\ref{eq:ZachReflXsec}) can be rewritten into the more commonly seen form~\cite{zachariasen1994theory,delRio_1992_Conceptual}:
\begin{equation}
    \gamma_0 \sigma(\alpha, \alpha_B) \approx Q G W(\alpha-\alpha_B - \Delta_M; \Gamma) \, .
    \label{eq:SimpleZachXsection}
\end{equation}
In comparison to Eq.~(\ref{eq:ZachReflXsec}), this form is quite accurate provided $\Lambda \ll \Gamma$, and is computationally much faster to evaluate.

However, as discussed previously, the crystallites are distributed on a unit sphere, and so the contribution from their distribution in $\psi$ needs to be accounted for. This also introduces an additional factor of $\cos(\alpha_B)$ due to the restriction of valid scattering geometries from the Laue equation~\cite{Bornemann_2020_Multiple, Wuttke_2014_Multiple}.
Furthermore, the polarization term needs to be modified to account for the difference in efficiency in the scattering for an angle of incidence $\alpha_B + \Delta$ on a crystallite oriented at $(\theta, \psi)$:
\begin{equation}
    G(\Delta, \phi-\psi) = 1 -  \sin^2(2[\alpha_B+\Delta])\sin^2(\phi-\psi) \, ,
    \label{eq:G_pol}
\end{equation}
and for an unpolarized beam:
\begin{equation}
    G_{\rm unpol}(\Delta) = \frac{1 +  \cos^2(2[\alpha_B+\Delta])}{2} \, ,
    \label{eq:G_unpol}
\end{equation}
The reflection cross-section should therefore be in the polarized case:
\begin{equation}
    \begin{split}
    \gamma_0 \sigma(\alpha, \alpha_B, \phi) &= Q \int_{-\alpha_B}^{\pi/2-\alpha_B} d\Delta \,  R(\Delta; \Lambda) \cos(\alpha_B+ \Delta) \\
    &\times \int_{-\pi}^{\pi} d\psi \, W(\theta(\alpha, \psi, \alpha_B+\Delta); \Gamma) G(\Delta, \phi-\psi)  \, ,
    \end{split}
    \label{eq:ReflXsectionPol}
\end{equation}
and in the unpolarized case:
\begin{equation}
    \begin{split}
        \gamma_0 \sigma_{\rm unpol}(\alpha, \alpha_B) &= Q  \int_{-\alpha_B}^{\pi/2-\alpha_B} d\Delta \, R(\Delta; \Lambda) \cos(\alpha_B+ \Delta)  G_{\rm unpol}(\Delta) \\
        &\times \int_{-\pi}^{\pi} d\psi \, W(\theta(\alpha, \psi, \alpha_B+\Delta); \Gamma)  \, ,
    \end{split}
    \label{eq:ReflXsection}
\end{equation}
where the $\Delta$ integration limits are now restricted to their physical limits.
Due to how narrow the IRC is, these integrals are computationally very expensive (in the context of the ray tracing) to evaluate numerically along $\Delta$. However, as with Eqs.~(\ref{eq:ZachReflXsec}) and~(\ref{eq:SimpleZachXsection}), provided that $\Lambda \ll \Gamma$, these integrals are well-approximated by:
\begin{equation}
    \begin{split}
        \gamma_0 \sigma(\alpha, \alpha_B ,\phi) &\approx Q \cos(\alpha_B+ \Delta_M) \\
        &\times\int_{-\pi}^{\pi} d\psi \, W(\theta(\alpha, \psi, \alpha_B+\Delta_M); \Gamma) G(\Delta_M, \phi-\psi)  \, ,
    \end{split}
    \label{eq:SimpleReflXsectionPol}
\end{equation}
\begin{equation}
    \begin{split}
        \gamma_0 \sigma_{\rm unpol}(\alpha, \alpha_B) &\approx Q G_{\rm unpol} \cos(\alpha_B+ \Delta_M) \\
        &\times \int_{-\pi}^{\pi} d\psi \, W(\theta(\alpha, \psi, \alpha_B+\Delta_M); \Gamma)  \, .
    \end{split}
    \label{eq:SimpleReflXsection}
\end{equation}
Due to the computational cost improvements and the fact that mosaic crystals used in spectrometers typically satisfy $\Lambda \ll \Gamma$, this approximation is used in \textit{HEART}.
If the unit sphere forms of the mosaic distribution functions are used, the integrals in $\psi$ are evaluated using trapezoid integration with varying step sizes in $\psi$: the integral is evaluated more densely around $\psi=0$ since the peak is very narrow and requires a denser grid of points to evaluate.

A comparison of the different mosaic functions and their reflectivities using Eqs.~(\ref{eq:SimpleZachXsection}) and~(\ref{eq:SimpleReflXsection}) is shown in Fig.~\ref{fig:Reflectivities}. The difference between the Gaussian and von Mises-Fisher distributions are minimal, both in terms of the distributions of the crystallites in $\theta$ and the reflectivity curve.
The reason for the similarity is that for small values of $\Gamma$ (e.g. in HOPG $\Gamma \sim 1^\circ$, so $\ll 1$~rad) the term $(\hat{n} - \hat{N})^2 \approx (\alpha - \alpha_B-\Delta_M)^2 + \cos^2(\alpha_B+ \Delta_M) \psi^2$~\cite{Bornemann_2020_Multiple, Wuttke_2014_Multiple}.
For the von Mises-Fisher $W_{\rm MF}$ from Eq.~(\ref{eq:MF}), integrating over $\psi$ in Eq.~(\ref{eq:SimpleReflXsection}) then recovers Eq.~(\ref{eq:SimpleZachXsection}), including the removal of the additional $\cos(\alpha_B + \Delta_M)$ term, when the angle $\alpha$ is within several $\Gamma$ of $\alpha_B + \Delta_M$~\cite{Bornemann_2020_Multiple}.

For the Lorentzian-based distributions, on the other hand, the reflectivity curve strongly depends on the choice of MDF. In terms of the angular distribution of the crystallites, the form of the wrapped Lorentzian $W_{\rm WL}$ we proposed in Eq.~(\ref{eq:WL}) produces a very similar distribution of the crystallites in $\theta$ as the Lorentzian. On the other hand, $W_{\rm SWL}$ produces a much narrower distribution in $\theta$ for the same value of $\Gamma$.
In terms of the reflectivity, $W_{\rm WL}$ produces a much flatter reflectivity curve than the Lorentzian, so reflections away from the Bragg angle will be able to substantially contribute relative to the peak, which will result in a more asymmetric instrument function~\cite{Gawne_2024_Effects}. Its peak reflectivity is also much lower than the Lorentzian, meaning the reflectivity of the crystal as a whole will be lower.
In contrast, $W_{\rm SWL}$ still has a very narrow reflectivity curve, leading to a more symmetric instrument function.

So far, the discussion has focused on just the scattering off the lattice planes $H=(hkl)$, which have a normal aligned along $\hat{n}$. However, there are also the antiparallel lattice planes $\overline{H}=(-h,-k,-l)$ which have a normal along $-\hat{n}$~\cite{zachariasen1994theory,Wuttke_2014_Multiple, Bornemann_2020_Multiple}.
Therefore, the reflectivity should account for both the scattering off $H$ and $\overline{H}$:
\begin{equation}
    \sigma = \sigma_{H}(-\arcsin(\hat{k}\cdot\hat{N}), \alpha_B) + \sigma_{\overline{H}}(\arcsin(\hat{k}\cdot\hat{N}), \alpha_B) \, .
\end{equation}
In general, the reflectivity of $H$ and $\overline{H}$ is the same since most crystals (in particular, the ones used in x-ray spectroscopy) have an inversion point. If this is not true, then $\sigma_{H} \neq \sigma_{\overline{H}}$. To account for this latter case, the code calculates $F_H$ and $F_{\overline{H}}$, so both terms in this sum can be calculated.
If a reflection does occur, $H$ or $\overline{H}$ is selected at random from the relative contribution of $\sigma_{H}$ and $\sigma_{\overline{H}}$ to $\sigma$. However, clearly the large angles involved in flipping the crystallite means that, unless the mosaicity is especially large, one term tends to dominate over the other depending on whether $\arcsin(\hat{k}\cdot\hat{N})$ is positive or negative.

Another assumption that has been made in the current reflectivity model is that the direction of the crystal normal $\hat{N}$ is constant along the direction of the photon. While this is true in the case of a flat crystal, many crystals in use have cylindrical or spherical geometry, meaning the crystal normal changes orientation as the photon travels through the crystal.
As a result, the angle of incidence $\alpha$ is changing as the ray passes through the crystal, and so the reflection cross-section should change too. To handle this, \textit{HEART} can optionally evaluate the normals at a number of points along the potential ray path -- defined by the user selecting the step size and maximum path length to consider -- and calculates the effective attenuation coefficient and the unnormalized Eq.~(\ref{eq:AnyEvent}) at each point.
From these, the CDF is calculated using the composite trapezoidal rule at each point along the ray path.
As the effective attenuation coefficient is typically very small, most of the rays will leave the crystal, requiring that the maximum distance be quite large to capture a sufficient amount of the tail outside the crystal in order to properly normalize the distribution.
At the same time, if the layers are used as a grid of possible photon steps for the ray, the step size would need to be very small to get a sufficiently continuous distribution. This would mean a huge number of a layers would need to have their normals and reflectivity evaluated at, which would imposes a large computational burden for any realistic simulation.
To account for the normal rotation more efficiently, a piecewise cubic Hermite interpolating polynomial interpolator~\cite{Fritsch_1984_Method} has been implemented in the code. 
To get the distance the ray travels, a random number is drawn between the maximum and minimum of the CDF, and the interpolator is used to find the associated distance.

A final addition to the code is the option to include scattering from higher orders of $H$; e.g. $H=(002)$, (004), etc., for graphite. When the order of the reflection is increased, the spacing between the lattice planes $d$ is proportionally reduced; e.g. $d_{(004)} = d_{(002)}/2$. From Eq.~(\ref{eq:BraggLaw}), this means that for the second order reflection (004), a higher energy photon $E$ can have the same Bragg angle as a photon with half the energy $E/2$ and can therefore be reflected to the same point on the detector.
This effect can be optionally accounted for in \textit{HEART} by defining multiple sets of Miller indices when setting up the crystal. The total reflectivity is then calculated by summing the contributions of the different $H$ together:
\begin{equation}
    \sigma = \sum_{H} (\sigma_{H} + \sigma_{\overline{H}}) \, .
\end{equation}
If a reflection occurs, the actual lattice plane the photon is reflected off is selected at random using the relative weights of the different reflectivities to $\sigma$. Since for each photon energy one reflection will tend to dominate, this approach should be sufficiently accurate for users who wish to account for this effect.

\subsubsection{Photon Path in the Crystal}

\begin{figure}
    \centering
    \includegraphics[width=\columnwidth,keepaspectratio]{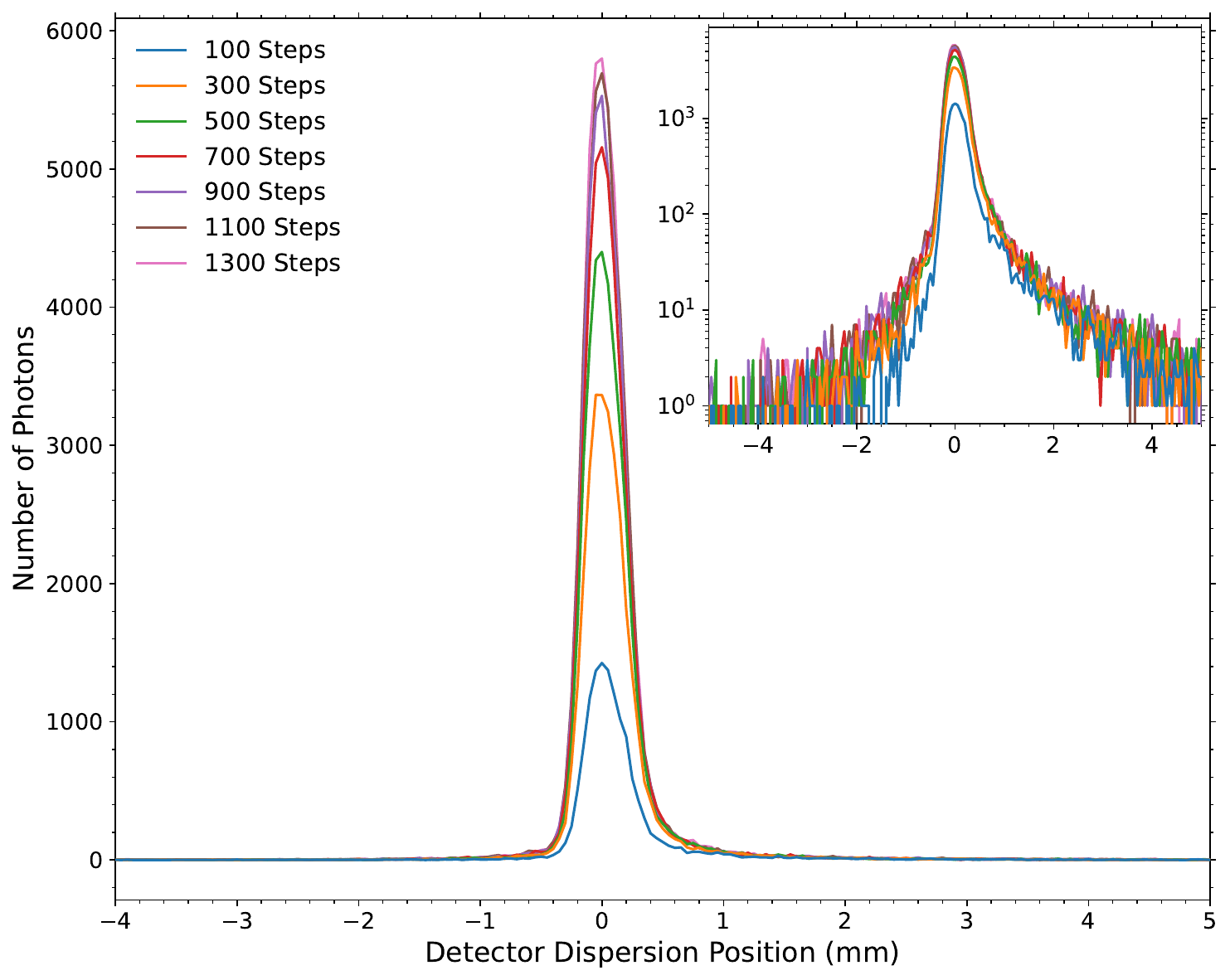}
    \caption{A plot showing the shape of the simulated instrument function for different burn-in periods. The inset shows the same spectra plotted on a log scale. The photon source is a point source of 10$^7$ 8.5~keV photons, incident on a von H\'amos spectrometer with a central photon energy of 8.5~keV. The crystal is graphite with a Lorentzian MDF and mosaicity of 0.14$^\circ$, and a Voigt IRC with 50.7~$\mu$rad and 10.0~$\mu$rad Gaussian and Lorentzian FWHMs, respectively. Also note the asymmetry of the instrument function, which is most apparent in the inset.
    }
    \label{fig:BurnIn}
\end{figure}

If the ray has entered the crystal, the distance $l$ it travels before an interaction  occurs is drawn at random from Eq.~(\ref{eq:AnyEvent}), with the effective attenuation coefficient $\mu$ calculated using the methods described in the Section~\ref{sec:ReflModel}.
The code then checks if this interaction point is still inside the crystal.
If it is not inside the crystal, the code then checks if the ray path will eventually intersect the detector and, if it does, it is added to the array of photons that will hit the detector. These rays are then no longer traced.

If the interaction point is inside the crystal, the ray is moved to the new position, and the type of interaction -- absorption or reflection -- is then determined: a random number is drawn from the uniform distribution $u[0,1)$, and the ray is absorbed if $u[0,1) \le \mu_0 / \mu_{\rm eff}$.
If the ray is absorbed, then it is removed from the simulation.

If instead the ray reflects, then the orientation of the crystallite needs to be determined.
This is done by sampling the integrands in either Eq.~(\ref{eq:ReflXsectionPol}), Eq.~(\ref{eq:ReflXsection}) or Eq.~(\ref{eq:ZachReflXsec}), depending on whether the user is considering polarization and if they are considering the wrapped forms of the distributions or not. If Eq.~(\ref{eq:ZachReflXsec}) is being used, the polarization terms follow the form of Eqs.~(\ref{eq:G_pol}) and~(\ref{eq:G_unpol}).
To sample these distributions, a Metropolis-within-Gibbs (MWG) sampler is used, with $\psi$ and $\Delta$ being alternately sampled.
The restrictions on the angle $\Delta \in [-\alpha_B, \pi/2 - \alpha_B]$ are enforced by rejecting angles which fall outside the range.
As the shape of the distribution depends on $\alpha$, $\alpha_B$ and, for polarization, $\phi$ -- all of which generally change between each ray (and reflections) -- and $\Gamma$, $\Lambda$, and $\Delta_M$ can also vary for each ray, the shape of the distribution change every step, necessitating a fresh burn-in period before drawing a sample on each ray step, with the length of burn-in period being an option for the user. A burn-in period of a few hundred steps is typically necessary, but the convergence of the spectrum versus burn-in should be checked as it can quite heavily affect results, as shown in Fig.~\ref{fig:BurnIn}.

Finally, the crystallite normal $\hat{n}_c$ is oriented according to the sampled $\Delta$ and $\psi$ by rotating the crystal normal $\hat{N}$ by an angle $\alpha - \alpha_B - \Delta$ about the axis $\hat{N} \times \hat{k}$, then rotating this vector about the ray by the angle $\psi$. The direction of the scattered ray $\hat{k}_s$ is given by the reflection equation
\begin{equation}
  \hat{k}_s = \hat{k} - 2(\hat{k} \cdot \hat{n}_c) \hat{n}_c \, ,
\end{equation}
and if polarization is being tracked in the crystal, this is updated to~\cite{zachariasen1994theory,Crowley_2014_Quantum}
\begin{equation}
    \hat{p}_s = \frac{\hat{p}_i - (\hat{p}_i \cdot \hat{k}_s) \hat{k}_s}{\sqrt{1 - (\hat{p}_i \cdot \hat{k}_s)^2}} \, .
\end{equation}

The above steps are repeated until a ray is either absorbed or it leaves the crystal.
This process repeats for every ray, with the number of rays in the simulation being the total number that will be traced, not the number that land on the detector.
Some useful additional data is tracked, such as the positions on the crystal that rays initially intersect, the number of rays that intersect, the number of rays absorbed in the crystal, and the number of rays that hit the detector.

\subsection{Detector}

The detector is simply a rectangle that the rays can hit, with bins defined along the two dimensions. Like the crystal, the detector can be arbitrarily oriented and positioned in space, allowing for arbitrary spectrometer geometries to be considered. At the end of the ray tracing simulation, the array of photons that hit the detector are binned into the pixels to produce two detector images: one being the number of photons in each bin, and the other being the total photon energy in each bin.
A simple treatment of the detector quantum efficiency is implemented, and explain in Section~\ref{sec:code}.
The physics of the detector beyond this very simple representation is not currently considered.


\section{Interacting with the Code}\label{sec:code}
\textit{HEART} has been built as a Python 3 package, so the typical use is to write a Python script or Jupyter notebook to setup simulation and run it. The primary way a user would interact with the code is by importing the \texttt{Spectrometer} class and using its methods to define the crystal, the detector, the shape of the source spot, and the photon energy profile. Finally, the ray tracing simulation is run by calling the class method \texttt{Spectrometer.ray\_trace()}, which takes some final inputs such as the number of photons and the number of burn-in steps. A step-by-step example is provided in the code documentation~\cite{HEART_docs}, alongside more detailed explanations of the code features than provided here. The code repository also contains a number of example scripts and a Jupyter notebook to show users how the code could be used.

The detector and crystal are also classes of their own stored within the \texttt{Spectrometer}. Their methods and properties can be accessed via \texttt{Spectrometer.detector} and \texttt{Spectrometer.crystal}, respectively. 
For the crystal, a user will need to use its methods to define properties of the crystal, such as its mosaicity and rocking curve.

To assist with calibrating the spectrometer if the dispersion on the detector is not known, an additional method \texttt{Spectrometer.calibrate\_spectrometer()} has been implemented. This runs the ray tracing simulation with the same parameters as the main ray tracer, but selects the photon energies at random from a uniform distribution. The limits of the distribution are defined by the input variables \texttt{min\_photon\_keV} and \texttt{max\_photon\_keV}. In the end, this produces an image of the energy dispersion on the detector, which a user can use to fit a dispersion curve.

At the end of the simulation, if the user has specified an output file, the data is saved into a HDF5 file.
Aside from the ray traced detector images and (if it has been calculated) calibration image, additional information on the spectrometer setup is also stored there. Optionally, the initial ray paths from the source to the crystal, the ray origins, and the individual photon energies of the rays can also be saved, although this is disabled by default since it results in much larger output files.
The data is also stored within the \texttt{Spectrometer.detector} object in case a user wants to directly interact with the data and not save a file. The important arrays are the photon hits on the detector (\texttt{detector.detector\_image}), the total photon energy image (\texttt{detector.detector\_image\_keV}), and the calibration energy map (\texttt{detector.calibration\_map}).

The \texttt{detector.detector\_image\_keV} array is the equivalent to the usual analogue-to-digital units (ADU) images that typically come from detectors in experiments since the ADU in a pixel is proportional to the total photon energy in that pixel. Indeed, some facilities automatically convert the ADU to keV for users based on detector calibrations. There is currently an optional, simple treatment for the quantum efficiency (QE) of the detector in \textit{HEART}. The user can provide an array of photon energies and QE values for their detector. For the rays that hit the detector, the photon energies are multiplied by their QE value (determined via linear interpolation) before being binned into the \texttt{detector.detector\_image\_keV} image.

As mentioned previously, the code is parallelised using threading and with MPI. Multithread parallelisation is done automatically, with the number of threads available being determined by Numba. However, the user can also choose the number of threads, and disable threading althogether, when calling the ray trace method.

For running a single simulation with MPI, the code should be initiated with \texttt{mpirun} -- an example Slurm submission script (\texttt{mpirun\_example.sh}) is provided in the examples directory and documentation. When a simulation is run like this, it should be noted that the complete detector images (and ray information if this was requested to be saved) after a simulation are only stored on the rank 0 processor. In testing the code on the NHR-NORD@G\"ottingen cluster, we found that the large arrays that needed to be communicated caused the \texttt{Allgather} and \texttt{Gather} functions from the mpi4py package~\cite{mpi4py} to hang. To circumvent this, we have implemented functions to do scatter and gather operations using the \texttt{Send} and \texttt{Recv} functions of mpi4py. However, since the arrays being handled are still large, sending these between large numbers of processors can take a long time. As a side effect of the manual implementation, rays can be unevenly distributed between processors, so the number of rays does need to be a multiple of the number of MPI processors.
It is also the rank 0 processor that handles saving data to the HDF5 file, which means H5py can be installed without MPI support (as is typically the case when installing from the package repositories).
A second Slurm script demonstrating how to submit multiple single simulations is also given the examples directory (\texttt{multirun\_example.sh}).

Additionally, since a number of facilities offer standardised spectrometer setups, some of these have been implemented in \textit{HEART} to allow a user to start running simulations with minimal setup. The use of these standard spectrometers is shown in the documentation and in the examples (the \texttt{standard\_X.py} scripts).

Otherwise, users are strongly encouraged to read the documentation~\cite{HEART_docs} and view the examples to see how to use the various code features.


\section{Ray Tracing Simulation Results}\label{sec:results}

In this section, we examine the ability of the ray tracer to produce spectra that are consistent with previous ray tracing results, and compare spectra produced by the ray tracer to experiment. We also examine differences between the different mosaic distribution and rocking curve functions available in the ray tracer.

\subsection{Simulated Detector Images}

\begin{figure*}
    \centering
    \includegraphics[width=\textwidth,keepaspectratio]{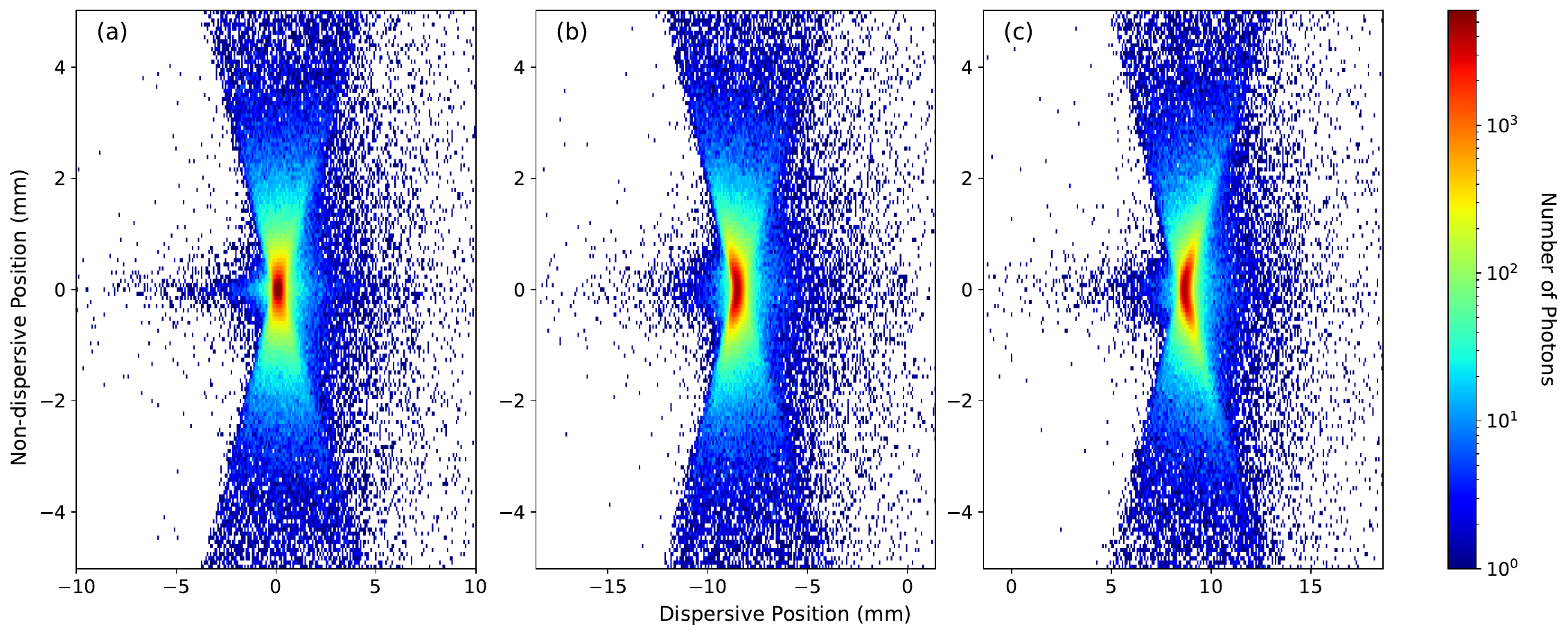}
    \caption{(a) A simulated detector image of a monochromated point source beam on a HAPG von H\'amos spectrometer that is in focus. (b) The same as (a), except the crystal is 1~mm above the optimal positioning for focusing. (c) The same as (a), except the crystal is 1~mm below the  optimal positioning for focusing.
    }
    \label{fig:VH_image}
\end{figure*}

\begin{figure}
    \centering
    \includegraphics[width=\columnwidth,keepaspectratio]{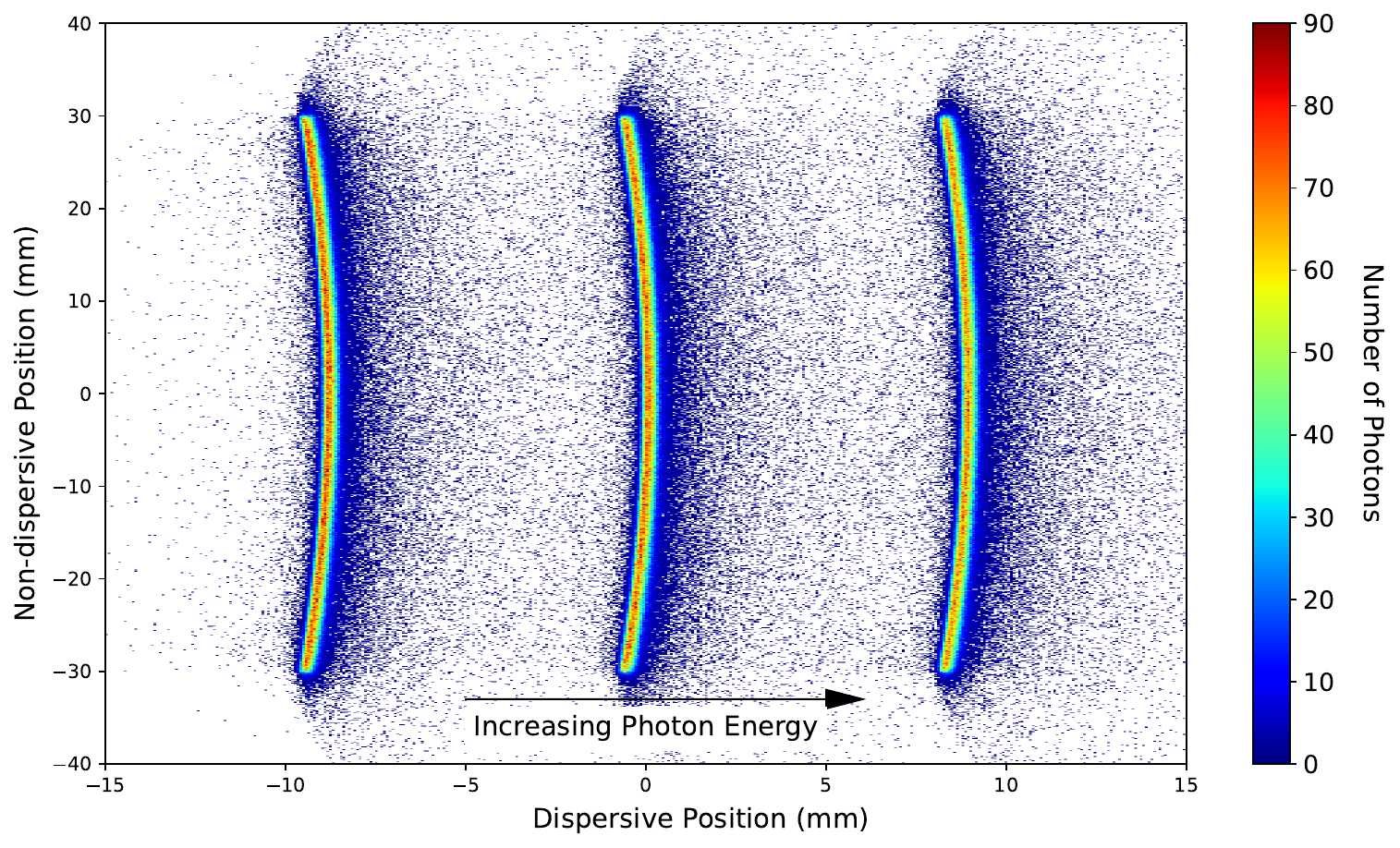}
    \caption{A simulated detector image of a flat HAPG crystal, incident with a point source emitting three photon energies (8.1, 8.2, and 8.3~keV). 
    }
    \label{fig:FC_image}
\end{figure}

\begin{figure}
    \centering
    \includegraphics[width=\columnwidth,keepaspectratio]{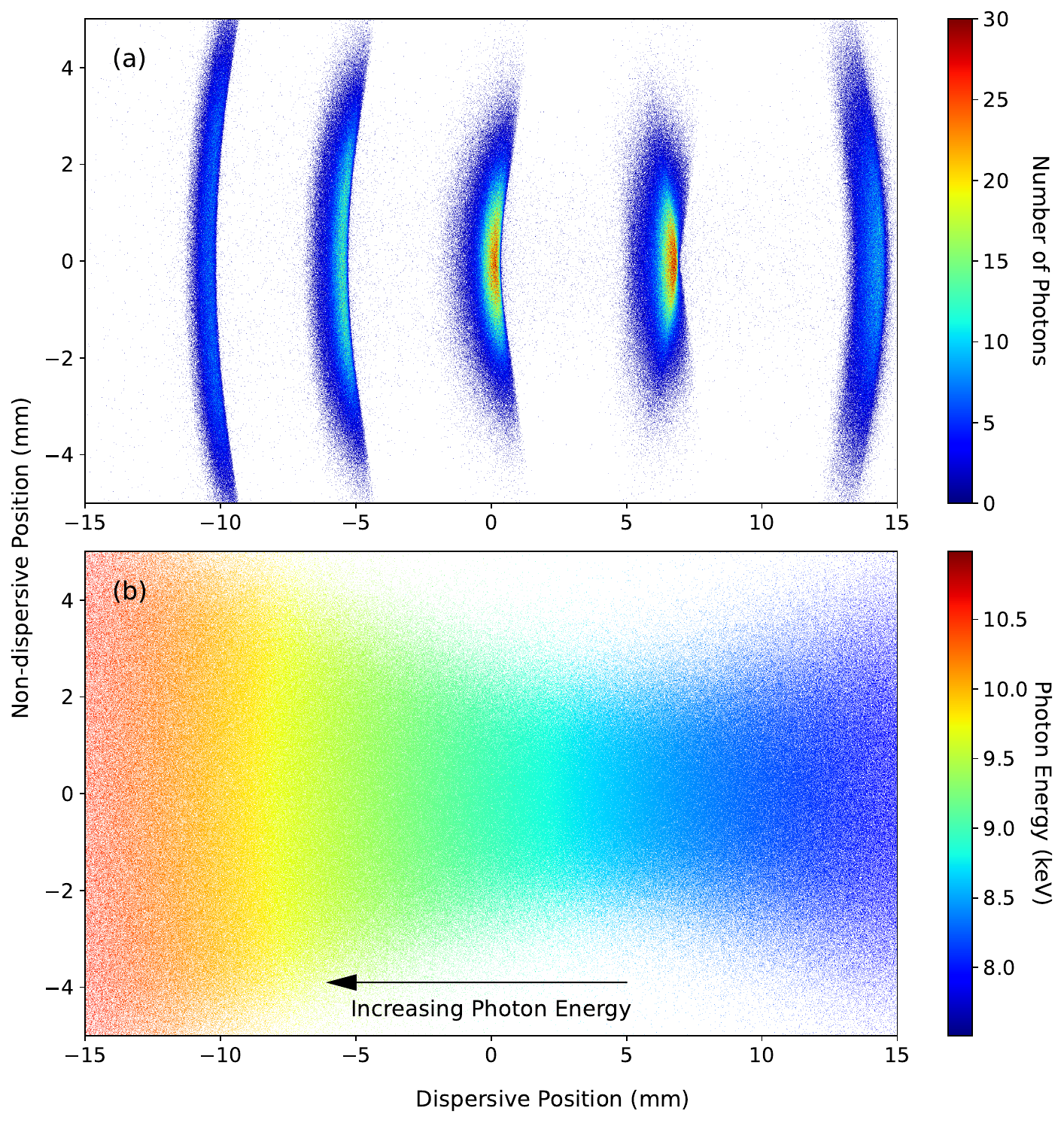}
    \caption{(a) A simulated detector image of the MACS spectrometer. The photon source is a point source emitting of five photon energies (8.0, 8.5, 9.0, 9.5, and 10.0~keV) with uniform intensity, and uniformly emitted in solid angle. (b) A ray traced image of the energy dispersion on the detector, showing the dispersive behaviour of the spectrometer, along with the focusing behaviour in the non-dispersive direction.
    }
    \label{fig:MACS_image}
\end{figure}

We start by examining whether the detector images produced by the ray tracer look reasonable.
As a first example, we look at a cylindrically bent HAPG crystal in von H\'amos geometry~\cite{vonHamos_1932}. Previous ray tracing studies of mosaic crystals in this geometry have shown that when in focus, a monochromated point source should resemble an `X' shape on the detector~\cite{2012_Zastrau_Focal}. This arises due to the mosaicity of the crystal, which causes focusing in the dispersive direction, but defocusing in the non-dispersive direction~\cite{delRio_1992_Conceptual}.
In Fig.~\ref{fig:VH_image}~(a), we show an image of a monochromatic 8.2~keV point source as viewed on a von H\'amos spectrometer using 40~$\mu$m thick HAPG crystal with a ROC of 80~mm that is in focus using the (002) reflection. We observe the expected `X'-shaped profile, with a well-focused bright spot at the centre of the detector.
When the crystal (or detector) is moved vertically off the best-focus positions, there should be additional defocusing as the detector now sits off the focal point of the mosaic focusing. In particular, a point source should resemble an arc rather than a small bright spot.
In Figs.~\ref{fig:VH_image}~(b) and~(c), the defocusing effect is shown for moving the crystal vertically by +1~mm and -1~mm from the optimal position, respectively.
As expected, the point source now resemble arcs rather than being a well-defined spot, and the direction of the curvature of the arcs depends on whether the crystal is above or below the optimal. The central position of the features have also shifted from the centre of the detector along the dispersion (horizontal) axis, which is again expected due to the change in the spectrometer geometry.

As a second example, we look at the shape of features on a flat crystal spectrometer. For a flat crystal, the detector image should consist of a set of circular arcs, with their ROC increasing with photon energy. In Fig.~\ref{fig:FC_image}, a ray traced image of three photon energies (8.1, 8.2, and 8.3~keV) incident on a flat crystal is shown. The simulation is identical as that in Fig.~\ref{fig:VH_image}, except the crystal is now flat, and the detector has been made wider to accommodate the larger features seen on the spectrometer. As expected, the three photon energies produce three arcs on the detector. We can also observe the edges of width of the crystal (30~mm) in the detector by the arcs having a length of 60~mm in the vertical direction, since the source-to-crystal and source-to-detector distances are equal.

As a final example, we look at the predictions of the ray tracer on the detector images of the MACS spectrometer at the NIF~\cite{Doeppner_2014_MACS}. This spectrometer uses a 300~$\mu$m thick cylindrically bent HOPG crystal using the (002) reflection, with a ROC of 54~mm in geometry that is similar to von H\'amos geometry, but with a key difference: while, the crystal is tilted so that the crystal's cylinder and dispersion axes are at ~$2.3^\circ$ to the diagnostic instrument manipulator (DIM) axis, the detector plane normal is aligned along the DIM axis. This in contrast to von H\'amos geometry where the detector plane lies along the cylindrical axes of the crystal to achieve focusing in the non-dispersive axis. From Ref.~\cite{Doeppner_2014_MACS}, we observe that the detector images from the spectrometer consist of large arcs that change shape along the central dispersion axis of the detector. Specifically, the arcs flip direction around the point on the detector associated with a photon energy around 8.6~keV, which is the focal point of the spectrometer. The size of the arcs increases at photon energies further away from the focal point since the detector position is more out of focus -- this functions in an identical manner to the defocusing behaviour in the von H\'amos geometry observed in Fig.~\ref{fig:VH_image}.

Ray traced detector images for the MACS spectrometer are plotted in Fig.~\ref{fig:MACS_image}~(a). The simulation has utilised a Gaussian MDF with a FWHM of $\Gamma = 1.3^\circ$ as determined in Ref.~\cite{Doeppner_2014_MACS}. The rocking curve of the crystallites is assumed to be the same as the previous simulations in this section, and is the same width for all photon energies. The photon source is taken to be a point source consisting of five photon energies (8.0, 8.5, 9.0, 9.5, and 10.0~keV), each with the same intensity, emitting uniformly in solid angle. $50\times10^6$ photons were sent to the spectrometer.
We observed similar shaped arcs across the detector as observed in Ref.~\cite{Doeppner_2014_MACS}, with their direction flipping around the focus at 8.6~keV, and their ROC increasing the further away from the focal point. The peak intensity of the features decreases further from the focal point, in part from the fact that photons are spread over larger arcs, but also from the reflectivity of the crystal for different photon energies and the solid angle coverage of the crystal.
All the arcs show a tail towards more negative directions on the horizontal detector axis (corresponding to higher photon energies), which is a result of the mosaic and depth broadening by the crystal.
The focusing and dispersive behaviour of the spectrometer is shown in Fig.~\ref{fig:MACS_image}~(b), which is the energy map of the detector, which is calculated by ray tracing photons sampled from a uniform distribution between 7.5--10.5~keV. The narrowest region on the vertical (non-dispersive) axis (i.e. best focus) is between $2.40$--5.40~mm along the horizontal axis, corresponding to photon energies around $8.5$--8.7~keV.
In general, we observe that the instrument function of the MACS spectrometer is strongly dependent on the photon energy, particularly due to the varying shapes and intensities of the different features along the spectrometer.

\subsection{Comparison with Experimental Data}

\begin{figure*}
    \centering
    \begin{subfigure}
        \centering
        \includegraphics[width=0.48\linewidth,keepaspectratio]{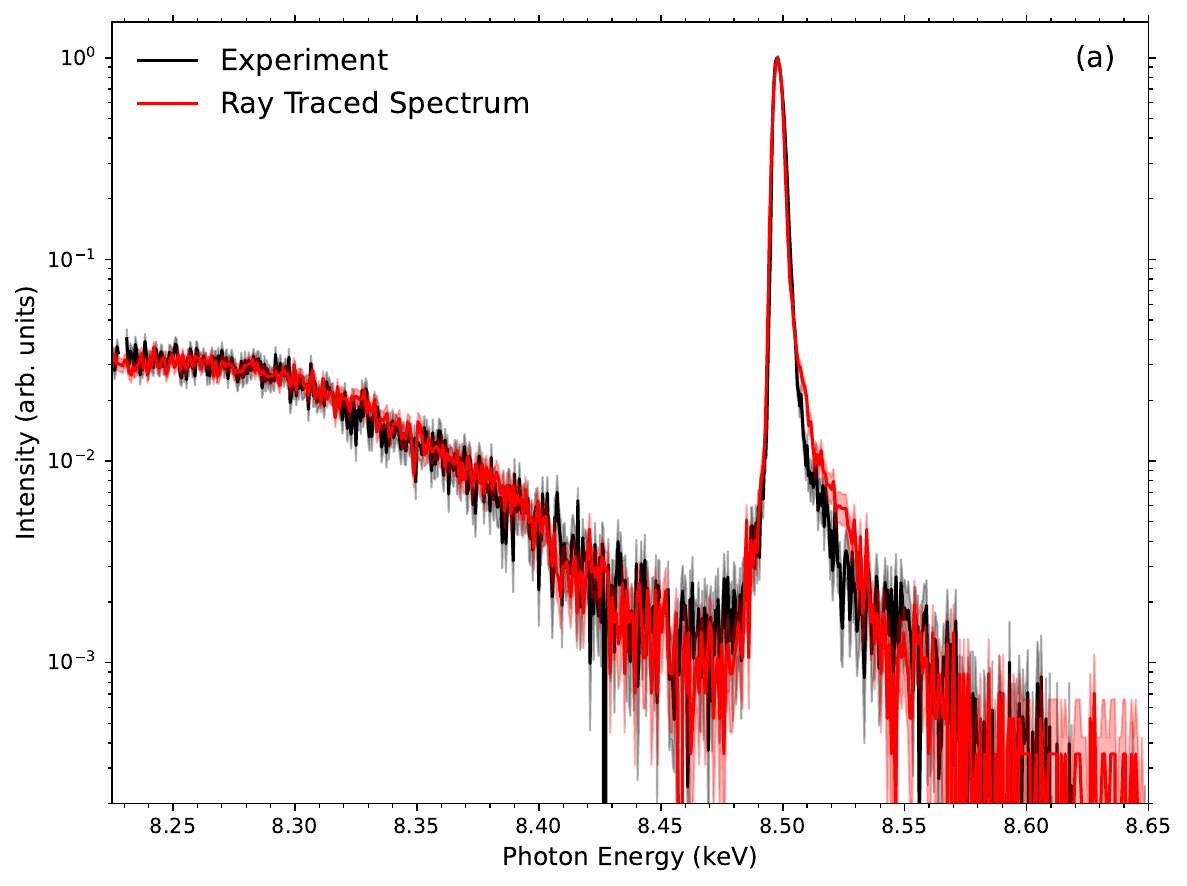}
    \end{subfigure}
    ~
    \begin{subfigure}
        \centering
        \includegraphics[width=0.48\linewidth,keepaspectratio]{XRTS_Comparison_345mm.pdf}
    \end{subfigure}
    \caption{(a) A comparison of a ray traced XRTS spectrum to one measured in experiment. The shaded areas represent the spectral uncertainty. The crystal is centred on the central photon energy as measured on the detector. (b) The same as (a), except the crystal is shifted closer to the source by 10~mm.}
    \label{fig:XRTS_Comp}
\end{figure*}

To evaluate the ability of the code to accurately model spectra, we now compare ray traced spectra to experimental data. The experimental spectra shown here were collected on the HAPG von H\'amos spectrometer~\cite{Preston_JoI_2020} at the HED Scientific Instrument at the European XFEL ~\cite{Zastrau_2021}. In both datasets, the crystal used was a 40~$\mu$m thick cylindrically bent HAPG crystal using the (002) reflection, with a radius of curvature of 80~mm. The size of the crystal is 80~mm in the dispersive direction, and 30~mm wide in the non-dispersive direction. The mosaicity and rocking curve of this specific crystal has not been directly measured before, meaning they are essentially free parameters to be determined in the modelling. These parameters were previously inferred in another study examining the instrument function of mosaic crystal~\cite{Gawne_2024_Effects}; however the approach used there is very different to this work, so we redetermine the parameters here.
Since the ray tracing simulations can (depending on the number of rays used) take up to a few minutes to run, and are inherently noisy in the spectra since we used numbers of rays that produced similar noise levels as the experimental data, we did not run a full least-squares optimization. Instead, for each spectrum a number of simulations were run, and the parameters that gave the best visual agreement were selected, which is sufficient for our purposes here. However, we emphasise that the code can be used as a part of better inference techniques, for example Bayesian estimators or Markov chain Monte Carlo methods, for fitting models to spectra.

We also include uncertainty bars on the simulated and experimental data, indicated by the shaded regions. For the experimental data, which is integrated over a number of frames, this is given by the frame-to-frame standard error. For the ray tracing simulations, the photon counting statistics are assumed to be Poisson distributed, so the standard deviation is calculated as $\sqrt{N_{\gamma}(E) + 1}$, where $N_{\gamma}(E)$ is the number of photons in the energy bin $E$.

In Fig.~\ref{fig:XRTS_Comp}~(a), we compare simulated and experimental XRTS spectra. The experimental data is from Ref.~\cite{Gawne_2024_Effects}, and is a scattering spectrum from ambient 8~$\mu$m thick polypropylene at a scattering angle of 166$^\circ$, measured with a monochromated 8.5~keV XFEL beam with a nominal 10~$\mu$m Gaussian spot on the target.
For the ray-traced spectrum, the quasi-elastic scattering is modelled as a delta function at 8.5~keV; and the inelastic feature (bound-free scattering off the electrons in the bonds) is modelled as a Gaussian centred at 8.25~keV with a FWHM of 0.257~keV, with a cut-off at an energy loss of 3~eV to treat the ionization threshold. The relative weights of the inelastic to elastic scattering were scaled to match the experiment (i.e. the ``ideal Rayleigh weight'').
The mosaic function used was Eq.~(\ref{eq:WL}) with a mosaicity of $\Gamma = 0.063^\circ$. The IRC was modelled as a Voigt profile with Gaussian and Lorentzian FWHM of 50~$\mu$rad and 8~$\mu$rad, respectively, giving a total FWHM of 54.3~$\mu$rad. A burn-in period of 1500 steps for the MWG sampler was found to be sufficient to converge the simulations.
As the detector has a central photon energy of $8.225$~keV, the crystal and detector were positioned at 345~mm and 690~mm along the dispersive direction in the simulation, respectively.
The simulation consisted of $20 \times 10^6$ photons emerging from the target, which produced similar noise levels as the experimental data. We choose this approach since it otherwise becomes harder to assess the agreement with the noisy high energy tail of the experiment.
The photon energies were randomly sampled from the input spectrum, and their origins determined by the shape of the XFEL spot and assuming uniform scattering in the target. The initial directions of the rays were selected to give uniform emission in solid angle.

Good agreement is seen between the experiment and the simulated XRTS spectrum, although the simulated spectrum overestimates the intensity of the high energy tail around $5.1$--5.3~keV.
However, this overestimation can be explained by one end of the crystal being observed. For mosaic crystals, the entire surface is reflective, and it can be shown that these reflections are always to more distant points on the detector (i.e. higher energy bins) compared to the nominal energy of the photon~\cite{Gawne_2024_Effects}. Once an edge of the crystal is reached, there is a drop in the reflectivity since only one side of the crystal now reflects~\cite{Gawne_2024_Effects}. Here, essentially all the measured photons above 8.5~keV come from the quasi-elastic feature, and at 8.53~keV there is a drop in the reflectivity due to one of the crystal edges along the dispersive direction being reached. Moving the crystal 10~mm closer to the source (but keeping the detector position fixed) moves the crystal edge position in the spectrum closer to the elastic scattering peak and results in better agreement with experiment at high energies, as shown in Fig.~\ref{fig:XRTS_Comp}~(b). The remainder of the spectrum is not noticeably changed by this repositioning.
This points to a slight subtlety in calibrations: when a spectrometer is calibrated using features with known energies (e.g. transition lines), what is really measured is just the total distance of the detector from the source. The crystal position is then usually assumed to be half this distance, but it is not guaranteed to be exactly correct -- the von H\'amos spectrometer used here is mounted on motors so that the total source-to-detector distance can be changed in order to fine-tune the measured spectral window. However, the crystal-to-detector distance is mechanically fixed upon installation, according to the requested central photon energy on the detector. Other features, such as the presence of crystal edges, may be useful in determining the exact position of the crystal {\it a posteriori}. This further highlights the need to make reliable instrument function measurements in experiment to determine the spectrometer geometry, particularly for analysis methods that are dependent on the specific shape of the instrument function.

This subtlety in crystal position was not previously considered. Von H\'amos geometry results in a one-to-two scaling from the crystal to the detector; i.e. for reflections from the surface of a perfect crystal, only 40~mm of the crystal (in the dispersive direction) would reflect on to the 80~mm detector used to measure the spectra. Indeed for this crystal, the length was deliberately chosen so that, without accounting for the mosaicity, the measured spectrum on an 80~mm long detector would be essentially constrained to a 40~mm section of the crystal in order to minimise edge effects~\cite{Preston_JoI_2020}. Nevertheless, the mosaic nature of the crystal means that reflections from the entire surface can be observed on the detector. The exact positions of the edges of the crystal therefore have an observable effect on the shape of the instrument function, which changes depending on where a photon has its highest reflectivity on the crystal~\cite{Gawne_2024_Effects}.

\begin{figure}
    \centering
    \includegraphics[width=\columnwidth,keepaspectratio]{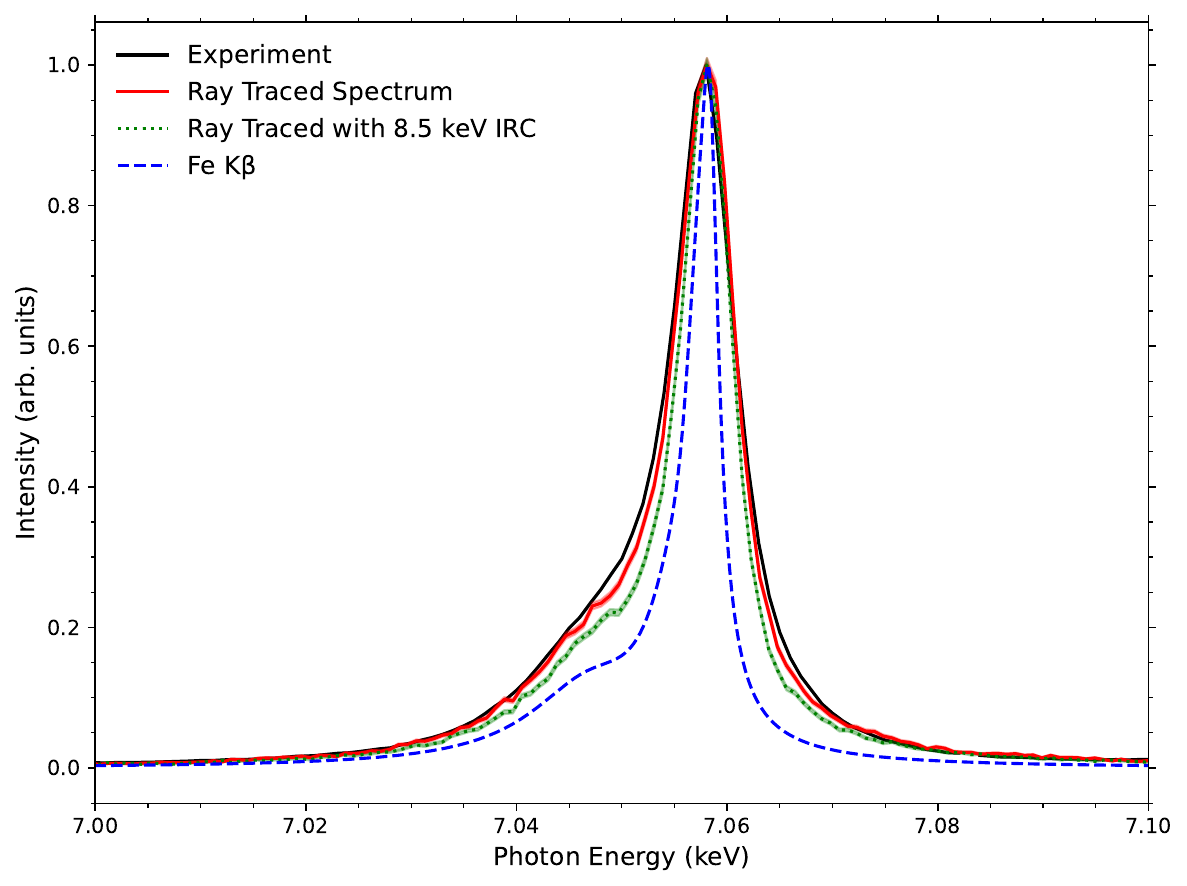}
    \caption{A comparison of the Fe $K_\beta$ spectrum from experiment (black) to a ray traced spectrum (red). The blue dashed line is the Fe $K_\beta$ lineshape from Ref.~\cite{Hoelzer_Lines}. The green dotted line shows the ray traced simulation using the same rocking curve parameters as the simulations in Fig.~\ref{fig:XRTS_Comp}. The shaded regions represent the spectral uncertainty in the ray traced and experimental spectra.}
    \label{fig:Kb_Comp}
\end{figure}

As a second example, in Fig.~\ref{fig:Kb_Comp} we compare spectra of the Fe $K_{\beta}$ lineshape, used as a calibrant in the experimental data from Ref.~\cite{Gawne_2024_Effects}. The target was a 3~$\mu$m thick Fe foil, and incident FEL had a nominal spot size of 10~$\mu$m. The XFEL was operating in seeded mode, with a self-seeded spike at 7702~eV sitting atop a roughly 20~eV wide self-amplified spontaneous emission (SASE) pedestal. Since the photon energy distribution of the XFEL is well above the Fe K-edge (7112~eV), and it should not affect the shape of the spectrum.
For the ray tracing simulation, the input Fe $K_{\beta}$ lineshape used was the multi-Lorentzian fit of H\"olzer \textit{et al.}~\cite{Hoelzer_Lines}, with $50\times10^6$ photon energies randomly sampled from the distribution. The crystal was positioned at the midpoint of the source and the central photon energy measured on the detector, i.e. the source-to-crystal distance and crystal-to-detector distance were equal.
We use the same mosaic function and width as the XRTS simulation, but now use a Voigt IRC with Gaussian and Lorentzian FWHM of 100~$\mu$rad and 20~$\mu$rad, respectively (total FWHM of 110.8~$\mu$rad). A burn-in period of 1500 steps was used for the MWG sampler. The shape of the source profile is accounted for in the same way as the XRTS data.

Again, we find good agreement between the experiment and the simulated XRTS spectrum, but there are visible differences between the two spectra. These might be reduced by running an optimization of the various crystal parameters (including its position), however this would be made more complicated by the aforementioned dependency of the shape of features on the position on the position of the crystal in space.
To get a reasonable looking fit, the IRC needed to be roughly twice as broad as that used for the XRTS simulation. Using the same IRC parameters as in Fig.~\ref{fig:XRTS_Comp} results in the simulated spectrum being too narrow, as shown in Fig.~\ref{fig:Kb_Comp}. This may be partially explained by the large difference in photon energies between the $K_\beta$ and XRTS spectra, and the fact that the IRC of a crystal is broader at lower photon energies~\cite{zachariasen1994theory,Gerlach_JAC_2015}. Furthermore, the Fe $K_{\beta}$ measured here was near the end of the detector and therefore far from the central photon energy, making it more susceptible to edge effects.
We were able to check that for different distributions of crystallite thicknesses, dynamic diffraction theory did predict the IRC would be wider for the lower photon energies, and we did find one distribution that had a factor of 1.5 difference. A more detailed investigation is hampered by the unknown distribution of crystallite thicknesses, but we conclude the difference between the two IRC widths used here is not unrealistic.
This highlights that even more care is needed in the simulation of spectrometers with large photon energy ranges (such as the MACS spectrometer) since the rocking curve width noticeably contributes to the broadening features, particularly in the shape of the low energy wing~\cite{Gawne_2024_Effects}.

\subsection{Comparison of Mosaic Distribution Functions}

\begin{figure}
    \centering
    \includegraphics[width=\columnwidth,keepaspectratio]{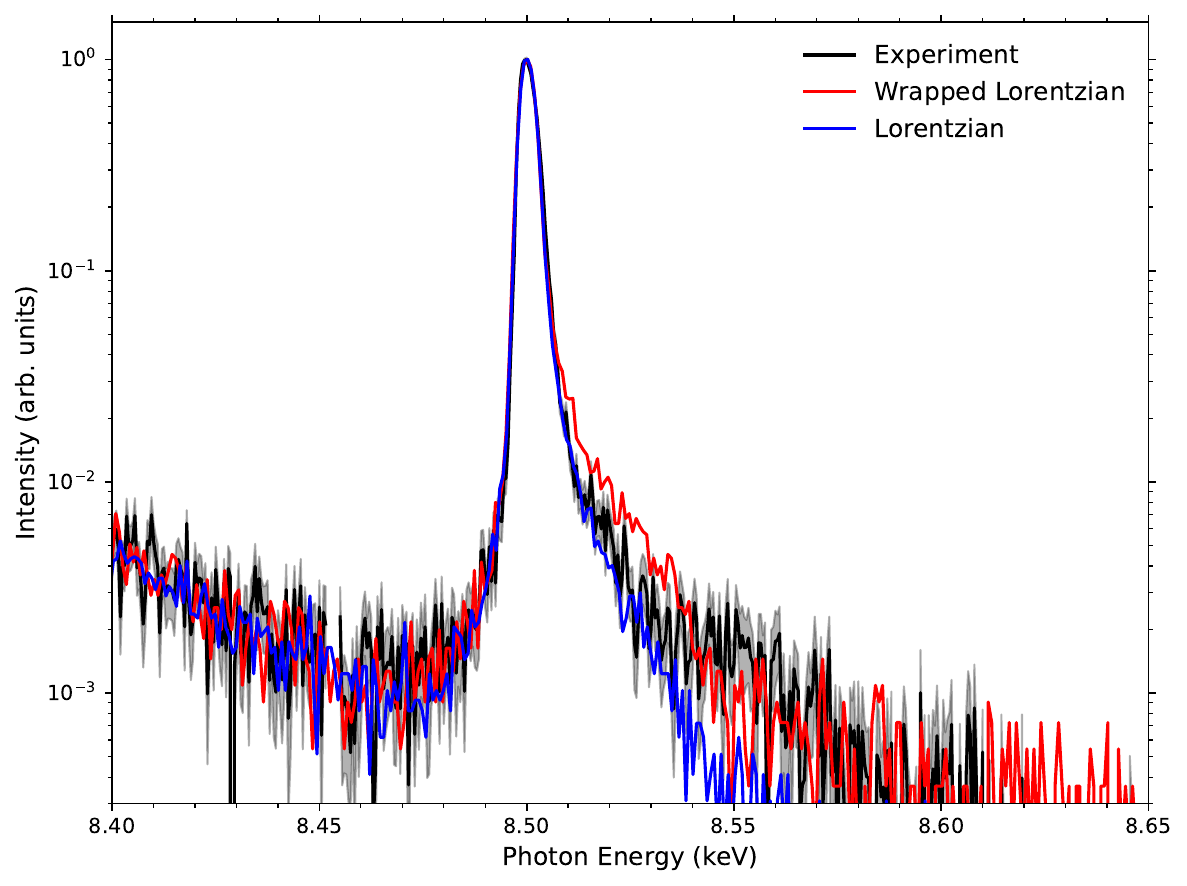}
    \caption{A comparison of the experimental (black) and simulated (red) XRTS spectra from Fig.~\ref{fig:XRTS_Comp}~(a) to an equivalent simulation using a Lorentzian MDF with mosaicity $\Gamma=0.15^\circ$ (blue).}
    \label{fig:WL_vs_L}
\end{figure}

\begin{figure}
    \centering
    \includegraphics[width=\columnwidth,keepaspectratio]{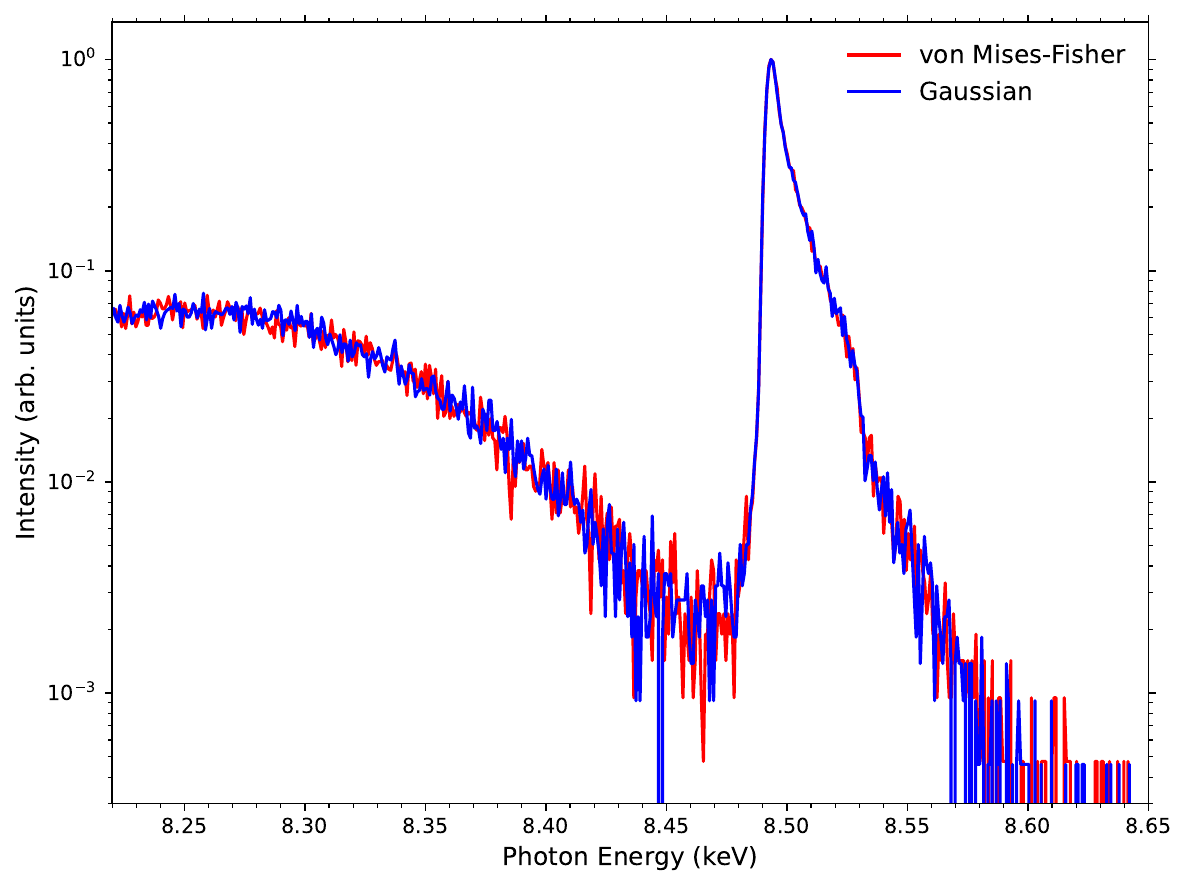}
    \caption{A comparison of simulated XRTS spectra using a von Mises-Fisher MDF from Eq.~(\ref{eq:MF}) (red) and a regular Gaussian (blue), both with mosaicity $\Gamma=1^\circ$.}
    \label{fig:MF_vs_G}
\end{figure}

We now compare results of ray traced simulations between the typically-used distributions and their wrapped equivalents.

In Fig.~\ref{fig:WL_vs_L}, we compare the experimental and ray traced simulations of Fig.~\ref{fig:XRTS_Comp}~(a) to a simulation using a regular Lorentzian distribution. The parameters are the same, except the Lorentzian simulation uses a larger mosaicty $\Gamma=0.15^\circ$. Utilising the same $\Gamma=0.063^\circ$ from earlier resulted in a spectrum that lacked a significant tail, so we have increased the mosaicity to make the results more comparable. Even with the much higher mosaicity, the Lorentzian distribution is still unable to reproduce the tail of the experimental data. This is because the tails of its reflectivity curve decay too quickly (see Fig.~\ref{fig:Reflectivities}~(d)), so the reflectivity is too concentrated around the nominal Bragg condition.
The discrepancies between the two simulated spectra compared to experiment might also suggest that an alternative distribution, which has a reflectivity curve between the Lorentzian and wrapped Lorentzian, may provide a better description of the crystallite distribution compared to what is used here.

In contrast, the simulated spectra using the von Mises-Fisher distribution from Eq.~(\ref{eq:MF}) and a regular Gaussian, both with mosaicity $\Gamma=1^\circ$, produce extremely similar spectra, as shown in Fig.~\ref{fig:MF_vs_G}.
Based on the reflectivity curves in Fig.~\ref{fig:Reflectivities}~(b), this result was to be expected.
This is ideal since it means a user can still get accurate simulations while benefiting from shorter simulation time by using the Gaussian distribution.

The simulated spectra in Fig.~\ref{fig:MF_vs_G} use the same input XRTS spectrum and crystal position as the Lorentzian case; only the form of the MDF has changed. The larger mosaicity used matches that for HOPG which typically has a larger mosaicity than HAPG. Here, the crystal edge is much more pronounced versus the Lorentzian case. Indeed, the quasi-elastic peak in this HOPG crystal shows a much more prominent asymmetry over the HAPG crystal in Fig.~\ref{fig:WL_vs_L}.

\subsection{Comparison of Rocking Curve Functions}

\begin{figure}
    \centering
    \includegraphics[width=\columnwidth,keepaspectratio]{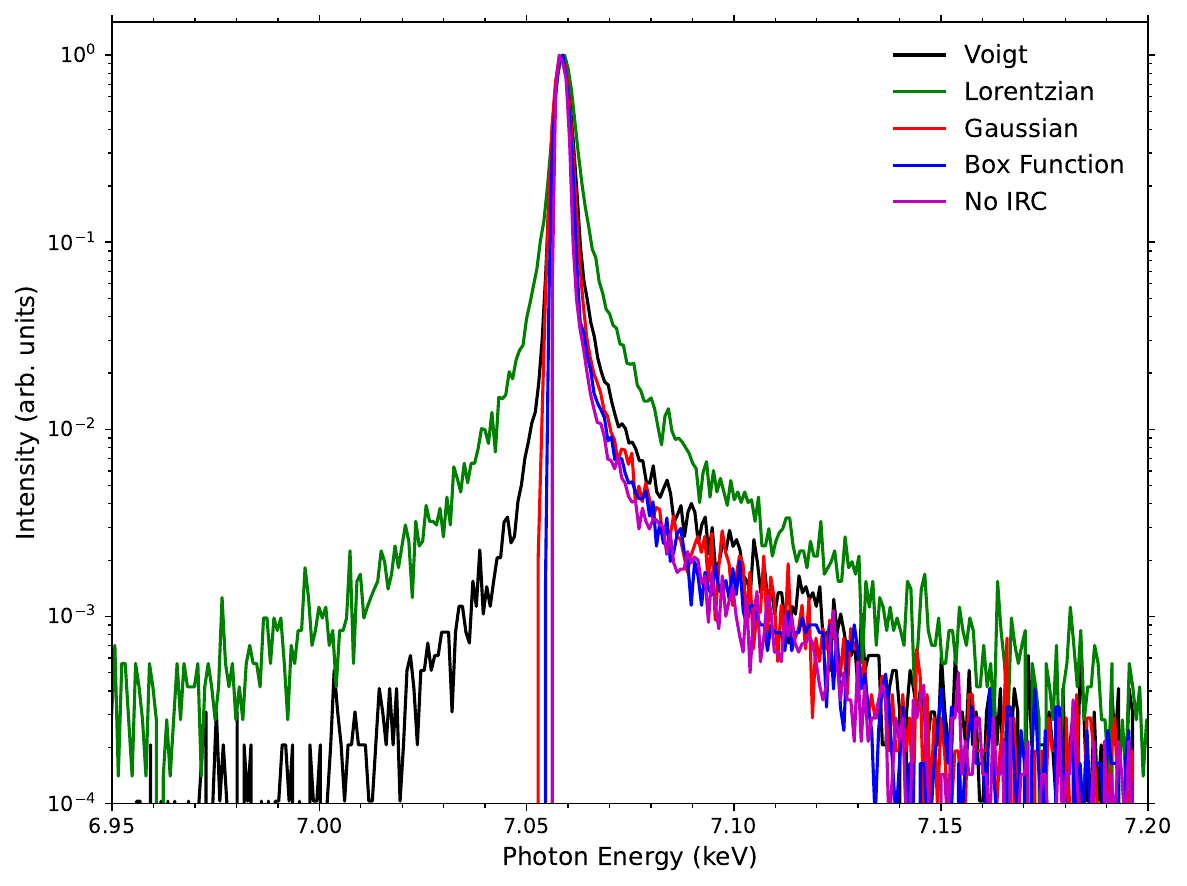}
    \caption{A comparison of simulated instrument function using different functions for the IRC, but with the same FWHM: a Voigt profile (black), a Lorentzian (green), a Gaussian (red), and a box function (blue). Also shown is the instrument function when neglecting the IRC entirely (magenta).}
    \label{fig:IRC_all}
\end{figure}

As a final example, and to demonstrate the necessity of using an IRC model with heavy tails, we compare the shape of the instrument function when using different forms of the IRC with the same FWHM -- a Voigt profile, a Gaussian, a Lorentzian, and a box function. We also include the shape of the instrument function when neglecting the IRC entirely.
For these simulations, we use the same crystal set up as that in Fig.~\ref{fig:Kb_Comp}, but now only use a monochromated beam of $7.06$~keV photons from a point source as the photon source in order to measure just the instrument function of the crystal.
For the Voigt profile, we estimated the Gaussian and Lorentzian FWHM to be 100~$\mu$rad and 20~$\mu$rad, respectively. This gives a total FWHM of 110.8~$\mu$rad, which is used in the Gaussian, Lorentzian and box function IRC calculations.

The resulting instrument functions are plotted in Fig.~\ref{fig:IRC_all}. The first observation is that only the pure Lorentzian and Voigt functions result in substantial broadening in the low energy wing, which has been observed in the experimental data here and in previous work~\cite{Gawne_2024_Effects}. This is expected from dynamic diffraction theory as the IRC of thin crystals is very extended~\cite{zachariasen1994theory}. However, the Lorentzian has extremely wide tails in both the low and high energy wings of the instrument function due to being a fat-tailed distribution. In general, we found a pure Lorentzian IRC struggles to fit the spectra, and in general it has a much narrower contribution to the IRC than the Gaussian contribution when using a Voigt IRC.

On the other hand, both the Gaussian and box function decay very quickly towards the low energy wing of the instrument function. The Gaussian is slightly broader than the box function on the low energy wing -- which is unsurprising given the distribution has tails while the box function does not -- but it is not sufficient to recover the low energy broadening seen in experiment. On the high energy wing, both the Gaussian and box function produce very similar looking spectra, since here the broadening is dominated by mosaic and depth effects. However, both distributions still produce instrument functions that are not as wide as the Voigt profile in this wing.

Finally, there is the case where the IRC is neglected. In this case, no broadening is seen in the low energy wing -- this is to be expected since mosaic and depth effects only broaden towards higher energies~\cite{Gawne_2024_Effects}. In the high energy wing, the instrument function looks very similar to the Gaussian and box function broadening, which demonstrates the dominance of the mosaic broadening over the IRC broadening in this wing if a thin-tailed distribution is used.


\section{Summary and Conclusions}\label{sec:conc}

We have introduced a new Monte Carlo ray tracing code, \textit{HEART}, which has been specifically designed to model  mosaic crystal spectrometers. The code is written in Python 3, and supports parallelization over multithreading and MPI.

We have demonstrated that the code is able to make good predictions versus experimental data in different layouts, and produces expected detector images for different experimental geometries.
Compared to another recent ray tracing code with mosaic crystal support~\cite{Smid_CPC_2021}, \textit{HEART} implements approximations enabling more detailed and more precise simulations, and is able to treat a wider class of mosaic distribution functions and rocking curves.

We have also incorporated into the code new distributions for the crystallites based on the fact that the crystallite normal vectors are distributed on the unit sphere~\cite{Bornemann_2020_Multiple,Wuttke_2014_Multiple}. As with previous work~\cite{Bornemann_2020_Multiple,Wuttke_2014_Multiple}, we find the Gaussian distribution and its unit sphere equivalent, the von Mises-Fisher distribution, produce very similar spectra. For the Lorentzian distribution, which is present in widely-used HAPG crystals, we propose a wrapped distribution for the crystallites that gives a similar distribution in the crystallite angle, but a much broader reflectivity curve. In comparison to experiment, we find the wrapped distribution produces better predictions than the Lorentzian distribution. However, both the standard Lorentzian and Gaussian, and their wrapped equivalents can be used in the code.

We also investigated the effect of different choices of the intrinsic rocking curves on the shape of the crystal instrument function, and highlight that only distributions with heavy tails can be used to approximate the broadening by the IRC of the crystallites~\cite{zachariasen1994theory,Gawne_2024_Effects}. However, for flexibility for the user, all the distributions presented here are available within the code.

We envisage that \textit{HEART} will be useful for a number of applications. For example, it can be useful for modelling actual experiments, such as in the planning stage by informing what sort of spectrometer setups would give the required resolution and photon statistics to observe desired features.
Additionally, the code may  also be used to perform realistic computer experiments, which may be useful for controlled theoretical studies of different analysis methods in-lieu of experimental data.
Furthermore, since the finite size effects of the target can be treated in the code, it may also be used in conjunction with e.g. radiative-hydrodynamic modelling to fully account for the finite size and density-temperature-ionization profile of a target. Such effects are expected to be important where target size and gradients are substantial, for example in NIF shots, and should be accounted for when assessing the validity of equation-of-state models and radiative-hydrodynamic simulations~\cite{Dornheim_2024_Wr}.
Another use case would be in modelling the instrument function of the spectrometer, which can be extremely difficult to accurately measure in experiments, but is nevertheless crucial for the correct interpretation of spectra~\cite{Dornheim_T_2022, Dornheim_T2_2022}.
As shown here and in a previous study~\cite{Gawne_2024_Effects}, the instrument function is found to depend on the photon energy and in the position of the spectrometer relative to the source. The instrument function therefore varies across the spectrometer, and simple approximations using convolutions may not be sufficiently accurate in some situations. For these cases, having a reliable way of predicting the spectrometer instrument function is all the more important for analysing spectral data.


\section*{Data Availability Statement}
The original experimental data for Fig.~\ref{fig:Kb_Comp} was reproduced with permission, and can be found here and is available upon reasonable request: doi:10.22003/XFEL.EU-DATA-003777-00.
The original experimental data for Figs.~\ref{fig:XRTS_Comp} and~\ref{fig:WL_vs_L} was reproduced with permission, and can be found here and is available upon reasonable request: doi:10.22003/XFEL.EU-DATA-005690-00.
The simulation data supporting in the manuscript is available at: doi:10.14278/rodare.3951.

\section*{Conflict of Interest Statement}
The authors have no conflicts of interest to disclose.

\section*{Acknowledgements}

This work was partially supported by the Center for Advanced Systems Understanding (CASUS), financed by Germany’s Federal Ministry of Education and Research (BMBF) and the Saxon state government out of the State budget approved by the Saxon State Parliament. 
This work has received funding from the European Union's Just Transition Fund (JTF) within the project \emph{R\"ontgenlaser-Optimierung der Laserfusion} (ROLF), contract number 5086999001, co-financed by the Saxon state government out of the State budget approved by the Saxon State Parliament.
This work has received funding from the European Research Council (ERC) under the European Union’s Horizon 2022 research and innovation programme
(Grant agreement No. 101076233, "PREXTREME"). 
Views and opinions expressed are however those of the authors only and do not necessarily reflect those of the European Union or the European Research Council Executive Agency. Neither the European Union nor the granting authority can be held responsible for them.

The authors gratefully acknowledge the computing time granted by the Resource Allocation Board and provided on the supercomputer Emmy/Grete at NHR-Nord@Göttingen as part of the NHR infrastructure. The calculations for this research were conducted with computing resources under the project mvp00024.

\bibliographystyle{elsarticle-num}
\bibliography{bibliography}

\appendix

\section{\label{sec:ExtraSampling}Sampling the Crystallite Orientation}

\begin{figure*}
    \centering
    \includegraphics[width=\textwidth,keepaspectratio]{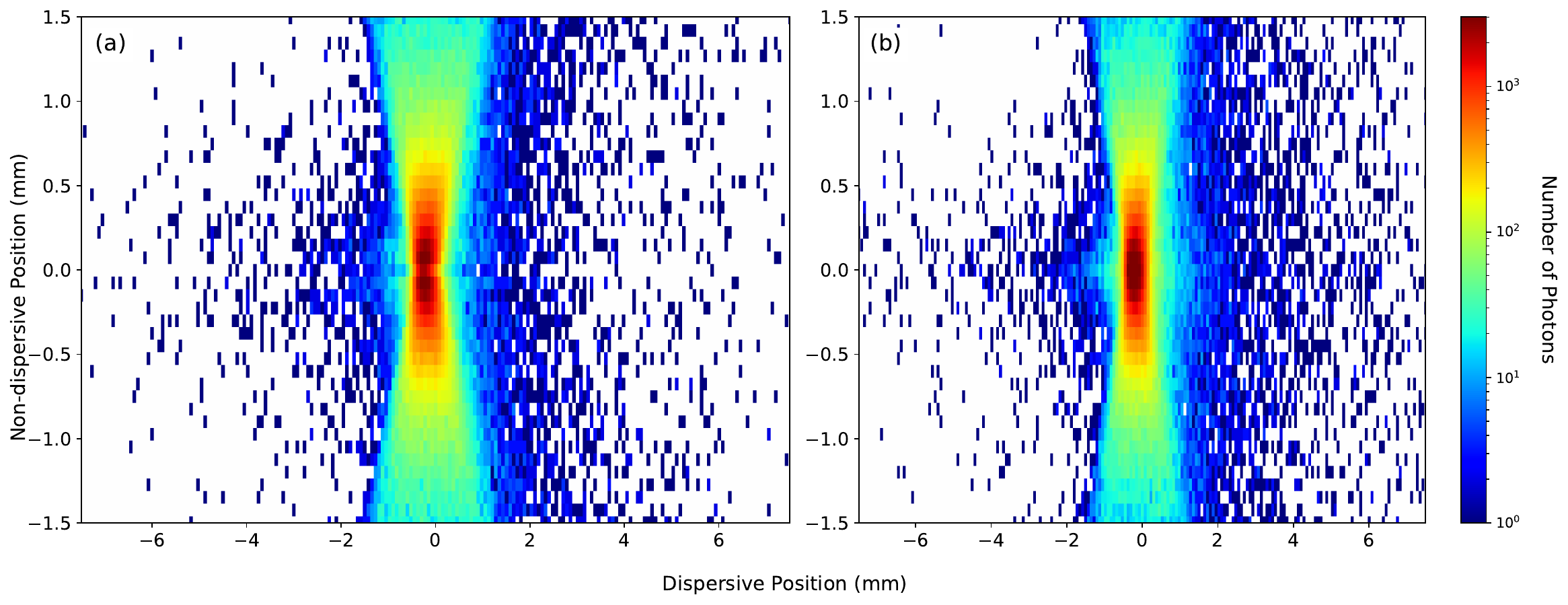}
    \caption{Ray traced detector images of 8.2~keV photons emerging uniformly from a point source, collected on a 40~$\mu$m thick HAPG (002) von H\'amos spectrometer, with a radius of curvature of 80~mm.
    (a) Sampling of the crystallite orientation is done by rejection sampling of the crystallite angle $\theta$, then inferring the angle $\psi$. Note the unphysical gap that appears at the centre of the detector image.
    (b) Sampling of the crystallite orientation is done by rejection sampling of the angle $\psi$, which also gives the crystallite angle $\theta$. Note the gap and double peaks are now replaced by a single spot.
    }
    \label{fig:CameraGap}
\end{figure*}

There is a slight subtlety in the correct way to determine the orientation of the crystallite that a photon reflects off. As mentioned in the main text, the reason the entirety of a mosaic crystal is reflective is because a photon has a probability of encountering a crystallite angled such that it can satisfy the Bragg condition. It seems natural then that in determining the crystallite angle, one only needs to sample the MDF to obtain the crystallite angle $\theta$ (within the limits of the allowed values), and the second angle $\psi$ can be inferred from this $\theta$ via Eq.~(\ref{eq:ThetaPsi}). Indeed, this is the approach described in Ref.~\cite{delRio_1992_Conceptual}.
However, this is approach is incorrect, and leads to an unphysical deficit in photon counts along the central dispersion axis. As an example, a ray traced spectrum on a von H\'amos spectrometer using this approach is shown in Fig.~\ref{fig:CameraGap}~(a), where the central axis clearly has fewer photons than the regions immediately surrounding it in the non-dispersive direction. Indeed, this sort of gap can be in Fig.~6 of Ref.~\cite{delRio_1992_Conceptual}, although they are harder to see than here. The reason for the deficit is due to an undersampling of angles around $\psi=0$.

The fundamental reason is because while the MDF does represent the PDF for finding a crystallite with an angle $\theta$ to the surface normal, it does not represent the PDF of a photon reflecting off a crystallite with a given orientation. The latter is given by the reflection cross-sections in Eqs.~(\ref{eq:SimpleReflXsectionPol}) and (\ref{eq:SimpleReflXsection}), which in the full three-dimensional cases involves an integral over the angle $\psi$. In other words, with the reflection cross-section representing the normalisation constant for a reflection PDF, the probability of a photon reflecting off a crystallite oriented between $[\psi-d\psi/2, \psi+d\psi/2]$ is $W(\theta(\alpha, \psi, \alpha_B+\Delta_M);\Gamma) d\psi$ in the unpolarised case (and $W(\theta(\alpha, \psi, \alpha_B+\Delta_M);\Gamma) G(\Delta_M, \phi-\psi) d\psi$ in the polarised case). The angle $\theta$ is then determined by the selected $\psi$ (and $\alpha$ and $\alpha_B+\Delta_M$) via Eq.~(\ref{eq:ThetaPsi}).
On the other hand, if one is purely interested in what is the probability of encountering a crystallite with an angle between [$\theta - d\theta/2$, $\theta + d\theta/2$] without any regard to reflections, this would be given by $W(\theta; \Gamma) d\theta$.
Once the reflectivity distribution is sampled, the detector images look as a expected with the central axis (i.e. the focal line) having the most photons, as shown in Fig.~\ref{fig:CameraGap}~(b).
We do note that this point on sampling $\psi$ rather than $\theta$ was more recently mentioned in part of the GitHub documentation for \emph{SHADOW3}~\cite{delRio_2013_Mosaic}.

Looking in more detail at the origin of the gap, we can look to what happens when $\theta$ is sampled and $\psi$ inferred. Using $W(\theta;\Gamma) d\theta$ as the probability of finding the crystallite within a given orientation, by a change of variable the probability of selecting the angle $\psi$ between $[\psi-d\psi/2, \psi+d\psi/2]$ is:
\begin{equation}
    \begin{split}
    W(\theta; \Gamma) d\theta = \sqrt{ \frac{C^2\sin^2(\psi)}{1 - [S+C\cos(\psi)]^2} } W(\theta(\alpha, \psi, \alpha_B); \Gamma) d\psi \, ,
    \end{split}
    \label{eq:Psi_From_Theta}
\end{equation}
where $S = \sin(\alpha)\sin(\alpha_B)$ and $C=\cos(\alpha)\cos(\alpha_B)$.
The $\psi$ PDF now has an additional factr compared to the reflection PDF. Aside from giving the incorrect reflectivity function, crucial to understanding the detector gap is that at $\psi=0$, the right hand side of Eq.~(\ref{eq:Psi_From_Theta}) is zero. In other words, around the angle $\psi =0$ which should maximise the reflectivity, sampling via  $\theta$ instead says these angles are not selected at all.

\begin{figure*}
    \centering
    \includegraphics[width=\textwidth,keepaspectratio]{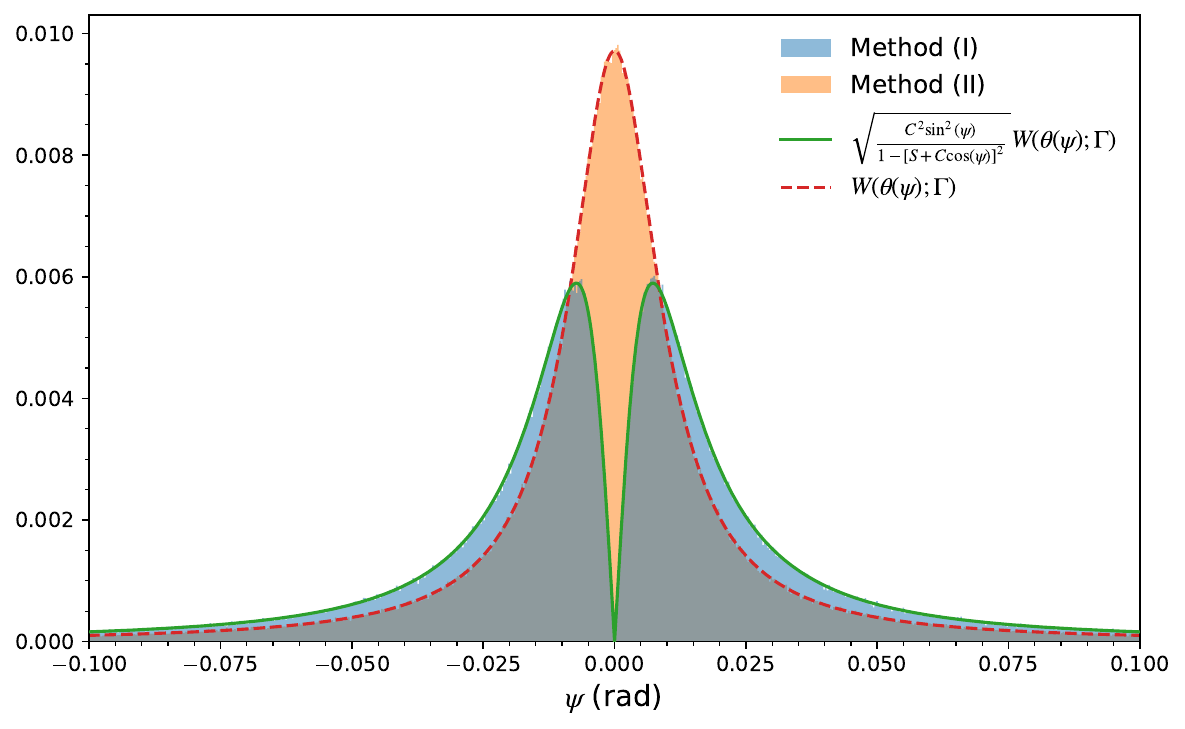}
    \caption{A comparison of sampling approaches the angle $\psi$ using two different distributions:    
    (I) Blue shaded area: Sampling the angle $\theta$ from the MDF within the range $\theta \in [|\alpha-\alpha_B|, |\pi - \alpha-\alpha_B|]$, and then inferring $\psi$ using Eq.~(\ref{eq:ThetaPsi}). This is sampling the probability of encountering a crystallite at a given orientation.
    (II) Orange shaded area: Sampling $\psi$ from a uniform distribution between $[-\pi, \pi]$, and performing an acceptance-rejection test with $W(\theta(\psi); \Gamma)$. Note that Method (II) is the correct sampling method for reflections as it samples the reflection cross-section distribution.
    The overlap between the two methods appears as a grey colour.
    The parameters used in these samplings are $\alpha_B = 0.2192$~rad, $\alpha - \alpha_B= 0.1$~rad, $\Gamma=0.06^\circ$, and $\sim10^6$ angles are sampled. The form of $W$ is used is a Lorentzian of the form Eq.~(\ref{eq:Lorentz}).
    The green-solid line indicates the PDF on the right-hand side of Eq.~(\ref{eq:Psi_From_Theta}), which matches the histogram from using Method (I). The red-dashed line indicates the PDF $W(\theta(\psi))$, which matches the histogram from Method (II).
    }
    \label{fig:MosaicDistributions}
\end{figure*}

A comparison is shown between the PDF of $\psi$ from $W(\theta(\psi); \Gamma)$, the right hand side of Eq.~(\ref{eq:Psi_From_Theta}), and histograms from sampling $\psi$ via the two distributions is shown in Fig.~\ref{fig:MosaicDistributions}. When sampling $\theta$ directly from the MDF, around $\psi=0$ there is a drop to zero in the number of counts and in the PDF -- this is the origin of the gap since these $\psi$ correspond to reflections along the dispersion axis. On both sides of the dip in the histogram and PDF there are maxima, which corresponds to the two peaks in intensity seen either side of the central line in Fig.~\ref{fig:CameraGap}~(a).

To summarise, when determining the crystallite orientation to determine the direction a photon should travel after reflection, there is a distinction between the probability of encountering a crystallite at a given orientation $(\theta, \psi)$ and the probability of reflecting off this crystallite.

\end{document}